\documentclass[twocolumn]{aastex701}
\usepackage{graphicx}	
\usepackage{amsmath}	

\usepackage{amssymb}	
\usepackage{bm}
\usepackage{amsfonts}
\usepackage{latexsym}
\usepackage{mathrsfs}
\usepackage{epsfig}
\usepackage{amsbsy}
\usepackage{enumitem}
\usepackage{multirow}
\usepackage[T1]{fontenc}

\begin{document}

\title{Radiation-Hydrodynamical Modeling of Accretion 
Columns in Ultraluminous X-ray Pulsars}

\correspondingauthor{Hong-Bo Li}
\email{lihb2020@pku.edu.cn}

\newcommand{\LZU}{Lanzhou Center for Theoretical Physics, Key Laboratory of Theoretical Physics of Gansu Province, Key Laboratory of Quantum Theory and Applications of MoE, Gansu Provincial Research Center for Basic Disciplines of Quantum Physics, Lanzhou 730000, China}

\newcommand{\LZUgravity}{Institute of Theoretical Physics \& Research Center of Gravitation, School of Physical Science and Technology, Lanzhou University, Lanzhou 730000, China}

\newcommand{\KIAA}{Kavli Institute for Astronomy and Astrophysics, 
           Peking University, Beijing 100871, China}

\newcommand{\PKUA}{Department of Astronomy, School of Physics, Peking University, Beijing 100871, China}

\newcommand{\IHEP}{State Key Laboratory of Particle Astrophysics, Institute of High Energy Physics, Chinese Academy of Sciences, Beijing 100049, China}

\newcommand{\UCAS}{University of Chinese Academy of Sciences, Chinese Academy of Sciences, 100049 Beijing, China}

\author[0000-0002-4850-8351]{Hong-Bo Li}
\affiliation{\LZU}
\affiliation{\LZUgravity}
\email{lihb2020@pku.edu.cn}

\author[0000-0002-8742-8397]{Ziming Wang}
\affiliation{\KIAA}
\affiliation{\PKUA}
\email{zwang@pku.edu.cn}

\author[0009-0008-0518-6795]{Bo-Yan Chen}
\affiliation{\IHEP}
\affiliation{\UCAS}
\email{bychen@ihep.ac.cn}

\author[0000-0003-3972-2564]{Shu Zhang}
\affiliation{\IHEP}
\email{szhang@ihep.ac.cn}

\author[0000-0002-9042-3044]{Ren-Xin Xu}
\affiliation{\KIAA}
\affiliation{\PKUA}
\email{r.x.xu@pku.edu.cn} 

\begin{abstract}

Ultraluminous X-ray pulsars (ULXPs) provide a unique laboratory for studying accretion-column physics under strongly magnetized super-Eddington conditions. In this work, we employ a one-dimensional radiation-hydrodynamical model to investigate the vertical structure of ULXP accretion columns, applying it to NGC 300 ULX-1, Swift J0243.6$+$6124, and SMC X-3.  We solve the coupled conservation equations for the inflowing gas and radiation field in a dipole-funnel geometry, allowing the velocity, density, gas pressure, radiation pressure, total energy transport rate, radiation sonic surface, and thermal mound location to be determined self-consistently. We find that radiation pressure naturally produces an extended sinking region above the stellar surface, where the inflowing gas is gradually decelerated and its kinetic energy is converted into radiation energy that partly escapes through the column walls. Despite their different luminosities, magnetic fields, and column geometries, the radiation sonic surfaces are tightly clustered near $15$ km, indicating a similar large-scale radiative deceleration structure. These results show that ULXP accretion columns develop extended radiation-dominated structures regulated by the mass accretion rate, magnetic scattering opacity, and column geometry. The dynamical profiles obtained here provide a physical basis for interpreting radiation-dominated accretion columns in ULXPs and for future physics-based spectral modeling.

\end{abstract}


\section{Introduction}
\label{sec: introduction}

Ultraluminous X-ray sources (ULXs) are extragalactic objects with X-ray luminosities $L_{\rm X}\gtrsim10^{39}\,{\rm erg\,s^{-1}}$, and in some cases up to $10^{42}\,{\rm erg\,s^{-1}}$, exceeding the Eddington limit for a stellar-mass black hole \citep{Kaaret:2017tcn, 2021AstBu..76....6F, King:2023nft}. They have traditionally been interpreted as accreting stellar-mass \citep{Begelman:2006bi, Poutanen:2006uc} or intermediate-mass black holes \citep{Colbert:1999es, 2006ASPC..352..121M, Maccarone:2007dd, Koliopanos:2017aja}. The discovery of coherent X-ray pulsations from M82 X-2 \citep{Bachetti:2014qsa}, with $L_{\rm X}\simeq1.8\times10^{40}\,{\rm erg\,s^{-1}}$ and a spin period of $1.37\,{\rm s}$, demonstrated that at least some ULXs are powered by accretion onto magnetized neutron stars. In these systems, known as ULX pulsars (ULXPs), the accreting gas is channeled by the magnetic field onto the magnetic poles, producing rotating hot spots and pulsed X-ray emission \citep{2010exru.book.....S}. The observed properties of ULXPs and candidate neutron-star ULXs are summarized in Table~\ref{tab: Table_1}.


Radiation-dominated accretion columns provide a natural framework for understanding the deceleration of the inflowing gas and the formation of the X-ray emission in bright accreting pulsars. In the classical picture, the radiation pressure gradient decelerates the supersonic flow above the stellar surface, producing an extended sinking region in which photons escape mainly through the column walls \citep{1976MNRAS.175..395B,1998ApJ...498..790B}. The transition from kinetic-energy-dominated inflow to radiation-dominated transport can be characterized by the radiation sonic surface. Near the base of the column, the thermal mound provides blackbody seed photons \citep{1973NPhS..246....1D} that are subsequently Comptonized in the accretion column. Its location is therefore part of the dynamical structure, rather than merely a spectral fitting parameter.

Multidimensional radiation-hydrodynamical simulations have provided important insight into the hydrodynamical behavior of supercritical column accretion. \citet{2016PASJ...68...83K} developed a two-dimensional radiation-hydrodynamical model for the accretion column of the ULXP M82 X-2. Their simulations showed that the inflowing gas is decelerated by a radiation-dominated shock above the stellar surface, forming an upper nearly free-fall region and a lower settling region. In this model, the kinetic energy of the accreting gas is converted into radiation near the shock, and the generated photons escape mainly through the column sidewall, allowing super-Eddington fan-beam emission. \citet{2020PASJ...72...15K} further considered the strongly magnetized case by suppressing gas motion perpendicular to the magnetic field and found that the accretion flow can form single- or multiple-hollow-cone structures depending on the mass accretion rate. These simulations highlight the importance of radiative shocks and lateral photon escape, but they were not designed to compute the two-fluid radial profiles required for spectral modeling, including separate electron and ion temperatures, Compton energy exchange, magnetic scattering optical depths, and the thermal mound location.

Models focusing on Comptonization in accretion columns have significantly improved the agreement between theoretical predictions and the observed phase-averaged X-ray spectra of several accretion-powered pulsars. In these models, the  emergent spectrum is formed through bulk and thermal Comptonization of bremsstrahlung, cyclotron, and blackbody seed photons \citep{2005ApJ...630..465B, 2005ApJ...621L..45B, 2007ApJ...654..435B, 2009A&A...498..509F, 2016A&A...591A..29F, 2021A&A...656A.105T, 2016ApJ...831..194W}. \citet{2022ApJ...939...67B} developed a generalized conical-geometry spectral model applicable to X-ray pulsars over a wide range of luminosities. However, such spectral models generally prescribe the velocity structure rather than solving for the full hydrodynamical profiles. A major step toward a self-consistent dynamical description was made by \citet{2017ApJ...835..129W, 2017ApJ...835..130W}, who constructed a two-fluid radiation-hydrodynamical model and applied it to the luminous X-ray pulsars Her X-1, Cen X-3, and LMC X-4. More recently, \citet{2024ApJ...966L...5W} applied a related framework to the ULXP RX J0209.6$-$7427 during its giant outburst. These studies demonstrate that self-consistent dynamical profiles are essential for connecting accretion-column physics with the formation of the observed X-ray spectra.

In this work, we implement the radiation-hydrodynamical framework to investigate the structure of accretion columns in ULXPs. Our goal is to determine the velocity, electron and ion temperatures, mass density, gas and radiation pressures, optical depths, total energy transport rate, radiation sonic surface, and thermal mound location from the coupled conservation equations, rather than prescribing these quantities in advance.  This is important in the ULXP regime, where the mass accretion rate and radiation pressure are much larger than in ordinary bright X-ray pulsars and may substantially modify the extent of the sinking region and the location of the thermalization layer. We apply the model to three ULXPs: NGC 300 ULX-1, Swift J0243.6$+$6124, and SMC X-3. These sources have measured or inferred luminosities in the ULXP regime and independent constraints on their magnetic fields from spin evolution, cyclotron resonance scattering features (CRSF), or luminosity-dependent pulse-profile transitions. The resulting self-consistent dynamical profiles provide the physical background required for subsequent physics-based spectral calculations.

The paper is organized as follows. In Section~\ref{sec: Phy_background}, we summarize the physical ingredients of the model, including magnetic scattering effects, the critical luminosity, and the equation of state. In Section~\ref{sec: hyd_eq}, we present the radiation-hydrodynamical conservation equations. In Section~\ref{sec: boundary_conditions}, we describe the boundary conditions at the column top, stellar surface, and thermal mound surface. In Section~\ref{sec: solv_equations}, we introduce the dimensionless formulation and the numerical solution procedure. In Section~\ref{sec: astro_apply}, we apply the model to NGC 300 ULX-1, Swift J0243.6$+$6124, and SMC X-3. We discuss the implications and limitations of our results in Section~\ref{sec: discussion}, and summarize the main conclusions in Section~\ref{sec: conclusion}.

\begin{table*}
    \renewcommand\arraystretch{1.5}
    \centering
    \caption{Observed properties of ULXPs and candidate neutron-star ULXs. The columns list the source name, spin period, period derivative, X-ray luminosity, distance, and reference.}
    \setlength{\tabcolsep}{0.05cm}
    {\begin{tabular}{l c c c c c}
    \hline
    \hline
    Source   & $P$   & $\dot{P}$   & $L_{\rm X}$   & $d$   & Ref.
    \\[-0.5em]
             & $(\rm s) $   & $(10^{-9}\, \rm s \,s^{-1})$   & $(10^{39}\,\rm erg\, s^{-1})$   & $(\rm Mpc)$   & \\
    \hline
    M82 X-2                & 1.37   & $-0.2$   & $18$        & 3.6   & [1] \\
    NGC 5907 ULX-1         & 1.13   & $-3.8$   & $\sim 100$  & 17.1  & [2] \\
    NGC 300  ULX-1         & 16.6   & $-556$   & $4.7$       & 2     & [3] \\
    NGC 1313 X-2           & 1.5    & $-13.8$  & $\sim 20$   & 4     & [4] \\
    NGC 2403 ULX           & 18     & $-100$   & $1.2$       & 3.2   & [5] \\   
    NGC 7793 P13           & 0.417   & $-0.04$  & $\sim 10$   & 3.6   & [6] \\
    NGC 7793 ULX-4         & 0.4    & $-35$    & $3.4$       &3.9    & [7] \\
    Swift J0243.6$+$6124 \footnote{For Swift J0243.6$+$6124, the distance $D=5.2\pm0.3\,{\rm kpc}$ is based on Gaia EDR3 \citep{2021AJ....161..147B, 2026ApJ...998..184C}. The corresponding peak luminosity, $L_{\rm X}\simeq 1.1\times10^{39}\,{\rm erg\,s^{-1}}$, is obtained by rescaling the original estimate of \citet{2018ApJ...863....9W} to the updated distance. Values in parentheses denote the earlier estimates adopted in our fiducial calculations.}   & 9.86   & $-22$    & $1.1~(2.0)$         & 0.0052 $\pm$ 0.003 & [8] \\
    SMC X-3                & 7.8    & $-6.46$  & $1.2$       & 0.062 & [9] \\
    RX J0209.6$-$7427      & 9.29   & $-17.5$  & $1.6$       & 0.06  & [10]\\
    M51 ULX-7              & 2.8    & $-0.15$  & $4$         & 8.6   & [11]\\
    NGC 4559 X7            & 2.6   & $-1$      & $5$         & 7.5  & [12]\\
    NGC 4631 X-8           & 9.66   & $-96$    & $3.4$       & 7.5  & [13]\\
    NGC 7456 ULX-1         & 4.55   & $-13$    & $\sim 40$   & 17.5 & [14]\\
    M51 ULX-8              & --    & --  & $2$              & 8.58   & [15] \\
    NGC 4656 ULX-1         & --    & -- & $32$              & 8.21  & [16]\\
    NGC 4861 ULX-2         & --    & -- & $\sim 3$          & 9.98  & [17]\\
    \hline
    \end{tabular}}
     \begin{list}{}{}
           \item $[{1}]$ \citet{Bachetti:2014qsa}; 
           \item $[{2}]$ \citet{Israel:2016chx, Fuerst:2023mrt}; 
           \item $[{3}]$ \citet{Carpano:2018pot, Vasilopoulos:2018vqh, Walton:2018ecy};
           \item $[{4}]$ \citet{Sathyaprakash:2019stu}; 
           \item $[{5}]$ \citet{Trudolyubov:2007nh};
           \item $[{6}]$ \citet{Motch:2013zth, Furst:2016mgk, Israel:2016sxx, Fuerst:2018ybc};  
           \item $[{7}]$ \citet{Quintin:2021beb}; 
           \item $[{8}]$ \citet{NICER:2018evw}; 
           \item $[{9}]$ \citet{Weng:2017qrc}; 
           \item $[{10}]$ \citet{Vasilopoulos:2020jtu};
           \item $[{11}]$ \citet{RodriguezCastillo:2019esm, Vasilopoulos:2019bhi};
          \item $[{12}]$ \citet{2025A&A...695A.238P};
          \item $[{13}]$ \citet{2025ApJ...994L..38D};
          \item $[{14}]$ \citet{2026arXiv260620308Y};
          \item $[{15}]$ \citet{Brightman:2018fsb, Bachetti:2021yxj};
          \item $[{16}]$ \citet{2026A&A...707L..20C};
          \item $[{17}]$ \citet{Allak:2026pqd}.
\end{list}
    \label{tab: Table_1}
\end{table*}

\section{Physical Background}
\label{sec: Phy_background}

\subsection{Magnetic Effects}
\label{sec: magnetic_effects}

In accretion-powered X-ray pulsars, the accretion flow is guided along the magnetic field lines by the strong magnetic field, typically with $B\sim10^{12}$--$10^{14}\,{\rm G}$. Such a strong field also affects the propagation of photons through the plasma. In particular, vacuum polarization makes the plasma birefringent, giving rise to two linearly polarized normal modes: the ordinary and extraordinary modes \citep{1979ApJ...228L..71C, 1979PhRvD..19.1684V, 1980ApJ...236..904N}. For the ordinary mode, the electric-field vector lies in the plane defined by the magnetic field and the photon propagation direction, whereas for the extraordinary mode, the electric-field vector is perpendicular to this plane.  The photon--electron scattering cross section therefore depends on the photon energy $\epsilon$ relative to the local cyclotron energy,
\begin{equation}
\epsilon_{\rm c}(r)
=
\frac{eB(r)h}{2\pi m_e c}
\simeq
11.57
\left[
\frac{B(r)}{10^{12}\,{\rm G}}
\right]
{\rm keV},
\label{eq: cyclotron_energy}
\end{equation}
as well as on the photon propagation direction and polarization state. Here $m_e$, $e$, $h$, and $c$ are the electron mass, the absolute value of the electron charge, Planck's constant, and the speed of light, respectively.

Throughout this work, $B_*$ denotes the magnetic-field strength at the stellar surface near the magnetic pole. Since the accretion flow is confined to a narrow dipolar funnel, we approximate the local magnetic-field strength along the column by the polar dipole relation
\begin{equation}
B(r)
=
B_*
\left(\frac{R_*}{r}\right)^3 .
\label{eq: dipole_B}
\end{equation}
This radial dependence is used when evaluating the cyclotron energy and cyclotron emission rate throughout the accretion column. The adopted surface field strengths $B_*$ are listed in Table~\ref{tab: model_parameters}.

A complete treatment of the energy- and angle-dependent scattering cross sections for ordinary- and extraordinary-mode photons is beyond the scope of the present work. We therefore follow \citet{1981A&A....93..255W, 2007ApJ...654..435B, 2022ApJ...939...67B} by adopting approximate scattering cross sections averaged over photon energy and over the two polarization modes. The cross sections for photons propagating parallel and perpendicular to the radial direction are denoted by $\sigma_{\parallel}$ and $\sigma_{\perp}$, respectively, and the angle-averaged cross section is denoted by $\bar{\sigma}$.

For photons propagating parallel to the magnetic field, the energy-averaged scattering cross section can be approximated as
\begin{equation}\label{eq: par_cross_section}
\sigma_{\parallel}
\simeq
\begin{cases}
\sigma_{\rm T}\,,
& \bar{\epsilon} \geq \epsilon_{\rm c}\,, \\[4pt]
\sigma_{\rm T}
\left( \bar{\epsilon}/\epsilon_{\rm c} \right)^2 \,,
& \bar{\epsilon} < \epsilon_{\rm c} \,,
\end{cases}
\end{equation}
where $\bar{\epsilon}$ is the radius-dependent mean photon energy. In practice, because the detailed photon spectrum is not solved in the present hydrodynamical calculation, we treat $\sigma_\parallel$ as an effective energy-averaged parameter. For photons propagating perpendicular to the magnetic field, we adopt the approximation \citep{1998ApJ...498..790B}
\begin{equation}\label{eq: per_cross_section}
\sigma_{\perp} = \sigma_{\rm T} \,.
\end{equation}

In this work, $\sigma_{\parallel}$ is treated as an effective model parameter that controls the coupling between radiation and matter along the column. Its adopted value is chosen so that the hydrodynamical solution satisfies the imposed boundary conditions for each source (see Table \ref{tab: model_parameters}). We also require the angle-averaged cross section $\bar{\sigma}$ used in the photon transport equation to satisfy
\begin{equation}\label{eq: aver_cross_section}
\sigma_{\parallel}
<
\bar{\sigma}
<
\sigma_{\perp} \,.
\end{equation}

\subsection{Critical Luminosity}
\label{sec: Cri_Luminosity}

Radiation pressure governs the dynamical structure of accretion flows in bright X-ray pulsars when the luminosity emitted above the magnetic pole exceeds the effective Eddington limit appropriate for the polar-cap geometry. This threshold is usually referred to as the critical luminosity and can be estimated as \citep{1976MNRAS.175..395B, 1991ApJ...367..575B, 1998ApJ...498..790B}
\begin{equation}\label{eq: crit_luminosity}
L_{\rm crit} = L_{\rm Edd} \frac{\pi r_0^2}{4\pi R_*^2}
\frac{\sigma_{\rm T}}{\sigma_{\parallel}} \,,
\end{equation}
where $r_0$ is the polar-cap radius, $R_*$ is the stellar radius, $\sigma_{\rm T}$ is the Thomson cross section, and $\sigma_{\parallel}$ is the electron scattering cross section for photons propagating parallel to the accretion-column axis. The quantity $L_{\rm Edd}$ is the standard spherical Eddington luminosity for pure, fully ionized hydrogen,
\begin{equation}\label{eq: Edd_luminosity}
L_{\rm Edd} =
\frac{4\pi G M_* m_p c}{\sigma_{\rm T}}
= 1.26 \times 10^{38}
\left(\frac{M_*}{M_\odot}\right)
\,{\rm erg\,s^{-1}} \,,
\end{equation}
where $M_*$ is the stellar mass, $m_p$ is the proton mass, and $c$ is the speed of light. The geometric factor $\pi r_0^2/(4\pi R_*^2)$ accounts for the reduction of the effective emitting area relative to the full stellar surface. The factor $\sigma_{\rm T}/\sigma_{\parallel}$ accounts for the increase in the critical luminosity in a strong magnetic field, where $\sigma_{\parallel}\lesssim\sigma_{\rm T}$ for photon energies well below the cyclotron energy \citep{1979PhRvD..19.3565M, 1979PhRvD..19.1684V}.

When the luminosity exceeds $L_{\rm crit}$, radiation pressure becomes strong enough to halt the freely falling gas above the neutron-star surface, and an accretion column develops above the magnetic polar cap. The inflowing gas then loses a significant fraction of its kinetic energy as it passes through an extended radiation-dominated shock. Within the shock region, photons are partly trapped by the converging flow and are advected downward toward the stellar surface, so that only a fraction of the generated radiation escapes locally. Downstream of the shock, the bulk velocity is strongly reduced, typically to $v\sim v_{\rm ff}/7$, and the flow enters a dense, slowly sinking region with high radiation energy density \citep{1976MNRAS.175..395B}. This sinking zone becomes the primary emitting region, with radiation escaping mainly through the column walls. 


\subsection{Equation of State}
\label{sec:EOS}

In ULXPs, the strong magnetic field can effectively confine the electron motion to the direction parallel to the magnetic field lines. As a result, the electron motion perpendicular to the field is quantized into Landau levels, while the parallel motion remains continuous. We therefore approximate the electron distribution as a one-dimensional Maxwellian along the magnetic field direction, with a mean thermal energy $\langle E_e\rangle=k_B T_e/2$, where $k_B$ is the Boltzmann constant. In contrast, the ion motion is not significantly affected by Landau quantization in the parameter regime considered here, and the ions are described by a three-dimensional Maxwellian distribution, with a mean thermal energy $\langle E_i\rangle=3k_B T_i/2$. The electron and ion internal energy densities are therefore
\begin{equation}\label{eq: internal_energy}
U_e = \frac{1}{2}n_e k_B T_e \,,
\qquad
U_i = \frac{3}{2}n_i k_B T_i \,,
\end{equation}
where $n_e$ and $n_i$ are the electron and ion number densities, respectively, and $T_e$ and $T_i$ are the corresponding temperatures.

The total pressure parallel to the local magnetic field is given by the sum of the electron, ion, and radiation components,
\begin{equation}\label{eq: eos_total_pressure}
P_{\rm tot} = P_e + P_i + P_{\rm rad}\,,
\end{equation}
where
\begin{equation}\label{eq: eos_gas_pressure}
P_e = n_e k_B T_e\,,
\qquad
P_i = n_i k_B T_i \,,
\end{equation}
are the electron and ion pressures, respectively. The radiation pressure, $P_{\rm rad}$, is not specified using a blackbody relation because the radiation field in X-ray pulsar accretion columns is generally nonthermal. Instead, it is determined self-consistently from the radiation energy conservation equation. The electron, ion, and radiation energy densities are related to their corresponding pressures through
\begin{equation}\label{eq: eos_energy_density}
\begin{aligned}
P_{e} &= \left(\gamma_e-1\right)U_{e} \,,
\\
P_{i} &= \left(\gamma_i-1\right)U_{i} \,,
\\
P_{\rm rad} &= \left(\gamma_{\rm rad}-1\right)U_{\rm rad}\,,
\end{aligned}
\end{equation}
where $\gamma_e=3$, $\gamma_i=5/3$, $\gamma_{\rm rad}=4/3$, and $U_{\rm rad}$ is the radiation energy density. These values reflect the one-dimensional electron motion, the three-dimensional ion motion, and the radiation equation of state, respectively.

The electron, ion, and radiation sound speeds, $a_e(r)$, $a_i(r)$, and $a_{\rm rad}(r)$ are defined by
\begin{equation}\label{eq: sound_speeds}
a_e^2 =
\frac{\gamma_e P_e}{\rho}\,,
\quad
a_i^2 =
\frac{\gamma_i P_i}{\rho}\,,
\quad
a_{\rm rad}^2 =
\frac{\gamma_{\rm rad}P_{\rm rad}}{\rho}\,.
\end{equation}
For a hydrogen plasma, we take $n_e\simeq n_i\simeq \rho/m_p$, where $\rho$ is the mass density, and $m_p$ is the proton mass. Combining this relation with Equation (\ref{eq: eos_gas_pressure}) gives
\begin{equation}\label{eq: gas_sound_speed_temperature}
a_e^2 =
\frac{\gamma_e k_B T_e}{m_p}\,,
\qquad
a_i^2 =
\frac{\gamma_i k_B T_i}{m_p}\,.
\end{equation}
There is no analogous temperature relation for $a_{\rm rad}$, because the radiation field is computed from the radiation energy equation rather than assumed to be thermal. Hence $a_{\rm rad}$ is determined self-consistently by the radiation pressure and mass density through Equation~(\ref{eq: sound_speeds}).

\section{The Hydrodynamic Equations}
\label{sec: hyd_eq}

In this section, we present a self-consistent calculation of the hydrodynamical structure of the accretion column in ULXPs. The dynamical structure is determined by five coupled quantities: the radiation sound speed $a_{\rm rad}(r)$, the electron sound speed $a_{e}(r)$, the ion sound speed $a_{i}(r)$, the bulk flow velocity $v(r)$, and the total energy transport rate $\dot{E}(r)$.

Following \citet{1982ApJ...257..733L}, the hydrodynamical evolution of the accretion flow is governed by the conservation of mass, momentum, and total energy. In the dipole-funnel geometry adopted here, these equations can be written as
\begin{equation}\label{eq: hyd_eq1}
\frac{\partial (\rho A)}{\partial t}
+
\frac{\partial (\rho v A)}{\partial r}
=0\,,
\end{equation}
\begin{equation}\label{eq: hyd_eq2}
\begin{split}
\frac{\partial (\rho v A)}{\partial t}
+
\frac{\partial (\rho v^2 A)}{\partial r}
+
A\frac{\partial}{\partial r}
\left(P_e+P_i+P_{\rm rad}\right)
\\
+
A\rho\frac{GM_*}{r^2}
=0\,,
\end{split}
\end{equation}
and
\begin{align}\label{eq: hyd_eq3}
\frac{\partial (U_{\rm tot}A)}{\partial t}
&+
\frac{\partial}{\partial r}
\left[
A\left(
\frac{1}{2}\rho v^3
+v(U_e+U_i)
+v(P_e+P_i)
\right.\right.
\nonumber\\
&\left.\left.
\qquad
+vU_{\rm rad}
+vP_{\rm rad}
-\kappa\frac{{\rm d}U_{\rm rad}}{{\rm d}r}
-\rho v\frac{GM_*}{r}
\right)
\right]
\nonumber\\
&=
A\dot U_{\rm esc}\,.
\end{align}
Here $\rho$ is the mass density, $v<0$ is the bulk velocity of the infalling gas, $U_{\rm tot}$ is the total energy density, and $A(r)$ is the cross-sectional area of the accretion column. The quantity $\kappa(r)$ is the spatial radiative diffusion coefficient along the accretion-column axis. Following \citet{2017ApJ...835..129W}, it is determined by the effective parallel electron-scattering cross section, $\sigma_\parallel$, and is given by
\begin{equation}
\kappa(r)
=
\frac{c}{3n_e(r)\sigma_\parallel}\,,
\label{eq:radiative_diffusion_coefficient}
\end{equation}
with units of ${\rm cm^2\,s^{-1}}$. The perpendicular scattering cross section is treated separately in the lateral photon-escape term.

In the dipole-funnel geometry, the solid angle occupied by the accretion flow varies with radius according to
\begin{equation}
\Omega(r)
=
\frac{r}{R_*}\Omega_*\,,
\end{equation}
where $\Omega_*$ is the solid angle subtended by the column footprint at the stellar surface. The corresponding cross-sectional area is therefore
\begin{equation}\label{eq: column_area}
A(r)
=
r^2\Omega(r)
=
\frac{\Omega_*r^3}{R_*}\,.
\end{equation}

The total energy density is defined as
\begin{equation}
U_{\rm tot}
=
\frac{1}{2}\rho v^2
+
U_e
+
U_i
+
U_{\rm rad}
-
\frac{GM_*\rho}{r}\,.
\end{equation}
The term $\dot U_{\rm esc}<0$ represents the net energy loss from the one-dimensional accretion column due to the lateral escape of photons through the column walls. Radiative emission and absorption, Compton scattering, and electron--ion Coulomb coupling redistribute energy internally among the electron, ion, and radiation components and therefore do not appear as net source terms in the total energy equation.

In a steady state, the mass continuity equation, Equation (\ref{eq: hyd_eq1}), becomes
\begin{equation}
\frac{1}{A}\frac{{\rm d}}{{\rm d}r}\left(A\rho v\right)=0 \,.
\end{equation}
Thus $A\rho v$ is conserved, and the mass accretion rate is related to
the density and velocity by
\begin{equation}\label{eq: mass_continuity}
\dot M
=
A(r)\rho |v|
=
\frac{ r^3}{R_*}\Omega_*\rho |v| \,.
\end{equation}
Equation~(\ref{eq: mass_continuity}) refers to the mass flow through the single magnetic funnel represented by our one-dimensional calculation. We do not evolve the accretion flows onto the two magnetic poles separately. Instead, we adopt an effective one-column approximation in which the modeled funnel represents the accretion-powered output of the source. 

If the total accretion flow were divided equally between two symmetric columns, the mass accretion rate through each column would be approximately half the total rate. Such a two-pole configuration is not explicitly considered in the present model.

The footprint solid angle is
\begin{equation}
\Omega_* =
2\pi\left(\cos\theta_1-\cos\theta_2\right) \,,
\end{equation}
where $\theta_1$ and $\theta_2$ denote the opening angles of the inner and outer walls of the column at the stellar surface, respectively. The corresponding arc lengths measured along the stellar surface, $\ell_1$ and $\ell_2$, are given by $\ell_1=\theta_1 R_*$ and $\ell_2=\theta_2 R_*$ (see Table~\ref{tab: model_parameters}). We also include the radius of the magnetic cap at the base of the accretion column, defined by $r_{0} \equiv R_{*} \theta_{2}$ \citep{2022ApJ...939...67B}. With the present notation, $r_0=\ell_2$. For a pure, fully ionized hydrogen plasma, charge neutrality gives $n_e=n_i$. Since $m_e\ll m_p$, the electron number density is
\begin{equation}\label{eq: electron_number_density}
n_e=n_i\simeq\frac{\rho}{m_p}
=
\frac{\dot M R_*}{\Omega_* r^3 |v| m_p} \,,
\end{equation}
where $n_e$ and $n_i$ are the electron and ion number densities, respectively.

\subsection{Radiation Sound Speed Derivative}
\label{sec: rad_speed}

In a steady state,  the total energy flux in the radial direction, averaged over the column cross-section at radius $r$, is given by
\begin{align}\label{eq: total_flux}
F(r) =
&\ \frac{1}{2}\rho v^3
+ v \left( P_{e} + P_{i} + P_{\rm rad}
\right.
\nonumber \\
&\left.
+ U_{e} + U_{i} + U_{\rm rad}
\right)
- \kappa \frac{{\rm d}U_{\rm rad}}{{\rm d}r}
- \rho v \frac{GM_*}{r} \,.
\end{align}
where the energy flux is defined to be negative for energy flow in the downward direction, and the accretion velocity is negative. The terms on the right-hand side of this equation represent the kinetic energy flux, the enthalpy flux, the radiation diffusion flux, and the gravitational energy flux, respectively. The total energy flux $F$ is related to the total energy transport rate in the radial direction, $\dot{E}$, via 
\begin{equation}\label{eq: energy_transport_rate}
 \dot{E}(r) = A(r) F(r) \,.
\end{equation}

By substituting the equation-of-state relations for the energy densities and pressures into Equation (\ref{eq: total_flux}), using Equation (\ref{eq: electron_number_density}) for the electron number density $n_e$, and replacing the total energy flux $F$ with Equation~(\ref{eq: energy_transport_rate}), we obtain the following first-order differential equation for the radiation sound speed $a_{\rm rad}$:
\begin{align}\label{eq: rad_speed}
\frac{{\rm d}a_{\rm rad}}{{\rm d}r}
&= \frac{a_{\rm rad}}{2v} \frac{{\rm d}v}{{\rm d}r}
+ \frac{3a_{\rm rad}}{2r} -
\frac{\sigma_{\parallel}\gamma_{\rm rad}\dot{M}}
{2m_p c\,a_{\rm rad}A(r)}\mathcal{E}(r) \,,
\end{align}
where 
\begin{equation}\label{eq: specific_energy_function}
\begin{split}
\mathcal{E}(r)
=\ & \frac{\dot{E}}{\dot{M}}
+ \frac{v^2}{2}
+ \frac{a_e^2}{\gamma_e-1}
+ \frac{a_i^2}{\gamma_i-1}
\\
& + \frac{a_{\rm rad}^2}{\gamma_{\rm rad}-1}
- \frac{GM_*}{r} \, .
\end{split}
\end{equation}

\subsection{Gas Sound Speed Derivative}
\label{sec: gas_speed}

Following \citet{2017ApJ...835..129W}, the energy equations for the  electrons and ions can be written as
\begin{align}\label{eq: conservation_gas}
\frac{{\rm D} U_e}{{\rm D}t}
&=
\gamma_e \frac{U_e}{\rho}
\frac{{\rm D}\rho}{{\rm D}t}
+ \dot{U}_{e} \,,
\\
\frac{{\rm D} U_i}{{\rm D}t}
&=
\gamma_i \frac{U_i}{\rho}
\frac{{\rm D}\rho}{{\rm D}t}
+ \dot{U}_{i} \,,
\end{align}
respectively, where the first terms on the right-hand side represent adiabatic compression, the final terms represent thermal coupling with the other species, and the comoving (Lagrangian) time derivative $D/Dt$ is defined by 
\begin{equation}\label{eq: Lagrangian}
\frac{{\rm D}}{{\rm D}t} \equiv \frac{\partial }{\partial t}
+ v \frac{\partial }{\partial r} \,.
\end{equation}
Here, the thermal coupling terms for the electrons include radiative heating and cooling, Compton energy exchange, and Coulomb coupling with the ions. We write
\begin{equation}\label{eq: processes}
\begin{aligned}
\dot{U}_{e}
&=
\dot{U}^{\rm emit}_{\rm brem}
+ \dot{U}^{\rm abs}_{\rm brem}
+ \dot{U}^{\rm emit}_{\rm cyc}
+ \dot{U}^{\rm abs}_{\rm cyc}
\\
&\quad
+ \dot{U}_{\rm Compton}
+ \dot{U}_{\rm ei}\,,
\\
\dot{U}_{i}
&=
-\dot{U}_{\rm ei}\,.
\end{aligned}
\end{equation}
Here, $\dot{U}^{\rm emit}_{\rm brem}$ and $\dot{U}^{\rm abs}_{\rm brem}$ denote bremsstrahlung emission and absorption, while $\dot{U}^{\rm emit}_{\rm cyc}$ and $\dot{U}^{\rm abs}_{\rm cyc}$ denote cyclotron emission and absorption. The term $\dot{U}_{\rm Compton}$ describes the energy exchange between photons and electrons through Compton scattering, and $\dot{U}_{\rm ei}$ represents the electron--ion Coulomb energy exchange \citep{1982ApJ...257..733L}. Expanding the comoving derivative in equation (\ref{eq: conservation_gas}), the result obtained in a steady state is
\begin{align}\label{eq: conservation_gas_radial}
\frac{{\rm d}U_e}{{\rm d}r}
&=
\gamma_e \frac{U_e}{\rho}
\frac{{\rm d}\rho}{{\rm d}r}
+ \frac{\dot{U}_e}{v}\,,
\\
\frac{{\rm d}U_i}{{\rm d}r}
&=
\gamma_i \frac{U_i}{\rho}
\frac{{\rm d}\rho}{{\rm d}r}
+ \frac{\dot{U}_i}{v}\,.
\end{align}
We can derive equivalent differential equations satisfied by the electron and ion sound speeds by using Equations (\ref{eq: eos_energy_density}), (\ref{eq: sound_speeds}), (\ref{eq: mass_continuity}) and (\ref{eq: conservation_gas_radial}) to obtain the following equation:
\begin{align}\label{eq: electron_speed}
\frac{{\rm d} a_{e}}{{\rm d}r} = - (\gamma_{e} -1) \left( \frac{a_{e}}{2 v} \frac{{\rm d} v}{{\rm d}r} + \frac{3a_{e}}{2r} \right)
\nonumber \\
 - \frac{1}{2}\gamma_{e}(\gamma_{e} -1) \frac{A(r)}{\dot{M}}\frac{\dot{U}_{e}}{a_{e}} \,,
\end{align}
and 
\begin{align}\label{eq: ion_speed}
\frac{{\rm d} a_{i}}{{\rm d}r} = -(\gamma_{i} -1) \left( \frac{a_{i}}{2 v} \frac{{\rm d} v}{{\rm d}r} + \frac{3a_{i}}{2r} \right)
\nonumber \\
 - \frac{1}{2}\gamma_{i}(\gamma_{i} -1) 
 \frac{A(r)}{\dot{M}}\frac{\dot{U}_{i}}{a_{i}} \,.
\end{align}

\subsection{Bulk Velocity Derivative}
\label{sec: bulk_speed}

The accreting gas is constrained by the strong magnetic field to move primarily along the field lines. Since there is no component of the Lorentz force parallel to the local magnetic field, the remaining acceleration in the parallel direction is due to the total pressure gradient and the gravitational field of the neutron star.  In the cross-section-averaged, one-dimensional approximation, the comoving acceleration along the radial direction can be expressed as \citep{1982ApJ...257..733L}
\begin{equation}
\frac{{\rm D} v }{{\rm D}t} = - \frac{1}{\rho}
\frac{\partial P_{\rm tot}}{\partial r} - \frac{G M_*}{r^2}  \,,
\end{equation}
where $P_{\rm tot} = P_{e} +  P_{i}+ P_{\rm rad}$ is the total pressure, and the Lagrangian time derivative $D/ Dt$ is defined by Equation (\ref{eq: Lagrangian}). Using Equations (\ref{eq: eos_energy_density}) and (\ref{eq: mass_continuity}) to substitute for the pressure components, $P_{\rm rad}$ and $P_{\rm gas}\equiv P_e+P_i$, and for the mass density, $\rho$, respectively, we obtain the following first-order differential equation for the bulk velocity $v$:
\begin{align}\label{eq: bulk_velocity}
\frac{{\rm d}v}{{\rm d}r} &=
\frac{v}{v^2- (a_{e}^2 +a_{i}^2)}
\left\{
\frac{3 (a_{e}^2 + a_{i}^2)}{r}
-\frac{G M_*}{r^2}
\right. \nonumber \\
&\quad \left.
+ \frac{\sigma_{\parallel}\dot{M}}{m_p c\, A(r)}
\mathcal{E}(r)
\right. \nonumber \\
&\quad \left.
+ \frac{A(r)}{\dot{M}}
\left[
\left(\gamma_{e}-1\right)\dot{U}_{e}
+ \left(\gamma_{i}-1\right)\dot{U}_{i}
\right]
\right\} \,.
\end{align}

\subsection{Total Energy Derivative}
\label{sec: total_energy_loss}

The total energy equation includes photon losses through the column walls. Although the gas is constrained to flow along the magnetic field lines, photons can diffuse laterally out of the column. In our one-dimensional, cross-section-averaged model, this leakage is treated using an escape-probability formalism. The total energy conservation equation is written as \citep{2017ApJ...835..129W}
\begin{eqnarray}\label{eq: total_energy_conservation}
\frac{\partial }{\partial t}\left( \frac{1}{2}\rho v^2
+ U_{e} + U_{i} + U_{\rm rad} - \frac{G M_* \rho}{r}
\right) & =  &\nonumber \\ 
- \frac{1}{A(r)}\frac{\partial }{\partial r}\left[A(r)F(r) \right]
+ \dot{U}_{\rm esc} \,,
\end{eqnarray}
where $F(r)=\dot E(r)/A(r)$ is the total energy flux. The volumetric energy loss rate due to photon escape is modeled as
\begin{equation}\label{eq: Uesc}
\dot{U}_{\rm esc}
= - \frac{U_{\rm rad} w_\perp}{\ell_{\rm esc}} \,.
\end{equation}
Here $w_\perp$ is the perpendicular diffusion velocity. Since this velocity cannot exceed the speed of light, we adopt the flux-limited form \citep{2017ApJ...835..129W}
\begin{equation}
w_\perp
= \min\left(c,\frac{c}{\tau_\perp}\right) \,,
\end{equation}
where the perpendicular scattering optical depth is evaluated locally as
\begin{equation}
\tau_\perp(r)
=
n_e(r)\sigma_\perp \ell_{\rm esc} \,,
\end{equation}
and we adopt $\sigma_\perp=\sigma_{\rm T}$ in this work. The corresponding photon escape distance is
\begin{equation}\label{eq: lesc}
\ell_{\rm esc}
=
(\ell_2-\ell_1)
\left(\frac{r}{R_*}\right)^{3/2} ,
\end{equation}
where $\ell_1$ and $\ell_2$ are the inner and outer arc-length radii of the column footprint,
\begin{equation}
\ell_1 = \theta_1 R_*\,,
\qquad
\ell_2 = \theta_2 R_* \,.
\end{equation}
In steady state, Equation (\ref{eq: total_energy_conservation}) reduces to
\begin{equation}\label{eq: Edot_steady}
\frac{1}{A(r)}
\frac{{\rm d}\dot E}{{\rm d}r}
=
- \frac{U_{\rm rad}}{\ell_{\rm esc}}
\min\left(c,\frac{c}{\tau_\perp}\right) .
\end{equation}
Using the column area $A(r)$ defined in Equation~(\ref{eq: column_area}), together with Equations~(\ref{eq: eos_energy_density}), (\ref{eq: sound_speeds}), and (\ref{eq: mass_continuity}), we obtain
\begin{equation}\label{eq: total_energy}
\frac{{\rm d}\dot E}{{\rm d}r} 
= - \frac{a_{\rm rad}^2 \dot M}{\gamma_{\rm rad}(\gamma_{\rm rad}-1)|v|\ell_{\rm esc}} \min\left(c,\frac{c}{\tau_\perp}\right) .
\end{equation}
This equation gives the radial attenuation of the total energy transport rate caused by lateral photon escape. Radiative emission, absorption, Compton scattering, and electron--ion energy exchange describe internal energy transfer among the gas and radiation components. These terms therefore cancel when the individual component energy equations are combined to form the total energy equation. Consequently, the only explicit energy-loss term retained in Equation~(\ref{eq: total_energy}) is the lateral photon-escape term $\dot U_{\rm esc}$.


\subsection{Energy Exchange Processes}
\label{sec: energy_exchange}

The energy-exchange rates per unit volume introduced in Equation~(\ref{eq: processes}) represent the radiative heating and cooling processes that couple the gas and radiation fields. In this subsection, we describe these processes: bremsstrahlung emission and absorption, cyclotron emission and absorption, Compton scattering, and electron--ion Coulomb energy exchange.

\subsubsection{Bremsstrahlung Emission and Absorption}
\label{sec: Bremsstrahlung_I}

Thermal bremsstrahlung emission is an important cooling channel for the ionized gas. During high-luminosity accretion onto pulsars, it also provides a dominant source of seed photons, which are subsequently Compton upscattered to produce the emergent X-ray spectrum \citep{2007ApJ...654..435B, 2017ApJ...835..130W, 2022ApJ...939...67B}. Assuming that the accreting gas is composed of fully ionized hydrogen, the thermal bremsstrahlung cooling rate per unit volume per unit frequency can be written as \citep[see][Equation (5.14)]{1979rpa..book.....R}
\begin{align}
\dot U_{{\rm brem},\nu}^{\rm emit}
={}&-
\left(
\frac{2^5\pi e^6}{3m_ec^3}
\right)
\left(
\frac{2\pi}{3k_{\rm B}m_e}
\right)^{1/2}
T_e^{-1/2} Z^2 n_e n_i
\nonumber\\
&\times
\exp\left(-\frac{h\nu}{k_{\rm B}T_e}\right)
\bar g_{\rm ff}\,.
\label{eq: brem_spectral}
\end{align}
Here $e$ is the absolute value of the electron charge in statcoulombs, $m_e$ is the electron mass, $k_B$ is the Boltzmann constant, $Z$ is the ion charge number, $\nu$ is the photon frequency, $h \nu$ is the photon energy, and $\bar{g}_{\rm ff}$ is the velocity-averaged Gaunt factor,  which accounts for quantum corrections.  For fully ionized hydrogen, we set $n_e=n_i$ and $Z=1$.  We further adopt $\bar{g}_{\rm ff}=1$, thereby neglecting quantum corrections.  The negative sign in Equation (\ref{eq: brem_spectral}) indicates that bremsstrahlung emission removes thermal energy from the electrons and therefore acts as a cooling term in the gas energy equation.  The total bremsstrahlung cooling rate over the photon-energy range adopted in our calculation is obtained by integrating Equation~(\ref{eq: brem_spectral}) over frequency. Using $d\nu=d\epsilon/h$, we obtain
\begin{align}\label{eq: ff_emission_rate_2}
\dot{U}_{\rm brem}^{\rm emit} &=
-\left(\frac{2^5 \pi e^6}{3 m_e c^3}\right)
\left(\frac{2\pi}{3 k_B m_e}\right)^{1/2}T_e^{-1/2} n_e^2 
\nonumber \\
&\quad \times
\frac{k_B T_e}{h}
\left[ \exp \left(-\frac{\epsilon_{\rm min}}{k_B T_e} \right) -
\exp \left(-\frac{\epsilon_{\rm max}}{k_B T_e} \right)
 \right] \,,
\end{align}
where $\epsilon_{\rm min}$ and  $\epsilon_{\rm max}$ are the minimum (lower) and maximum (upper) photon energies of the computational domain, respectively. In this work, we set these boundaries to $0.01\,{\rm keV}$ and $200\,{\rm keV}$.

For bremsstrahlung absorption, we follow \citet{2017ApJ...835..129W} and include the heating of electrons due to free--free absorption of low-frequency radiation in the accretion column. This process plays an important role in regulating the gas temperature. The frequency-integrated heating rate per unit volume due to thermal bremsstrahlung absorption is given by
\begin{equation}\label{eq: ff_emission_rate_3}
\dot{U}_{\rm brem}^{\rm abs}
=U_{\rm rad}\,\alpha_{\rm R}\,c \, ,
\end{equation}
where $U_{\rm rad} =P_{\rm rad}/(\gamma_{\rm rad} -1) $, and $\alpha_{\rm R}$ is the Rosseland mean absorption coefficient for fully ionized hydrogen. In cgs units, it can be written as \citep{1979rpa..book.....R}
\begin{equation}
\alpha_{\rm R}
= 1.7 \times 10^{-25} T_e^{-7/2} n_e^2 \,.
\end{equation}
In Equation (\ref{eq: ff_emission_rate_3}), the positive sign of $\dot{U}_{\rm brem}^{\rm abs}$ indicates that free--free absorption transfers energy from the radiation field to the electrons and therefore acts as a heating term in the gas energy equation.

\subsubsection{Cyclotron Emission and Absorption}
\label{sec: Cyclotron_I}

Electrons in the accretion column can also exchange energy with the radiation field through the emission and absorption of cyclotron radiation. Cyclotron photons are produced when electron--proton collisions excite electrons from the ground Landau level, $n=0$, to the first excited level, $n=1$. These excitations are followed rapidly by radiative de-excitation back to the ground state. Under the densities and temperatures typical of pulsar accretion columns, cyclotron absorption is therefore effectively followed by prompt radiative re-emission. As a result, the process can be treated as resonant scattering, producing no net change in the angle-averaged photon distribution \citep{1980ApJ...236..904N, 1987ApJ...312..666A}. Consequently, cyclotron absorption does not lead to net gas heating in our dynamical model, and we set $\dot{U}_{\rm cyc}^{\rm abs}=0$. Near the column surface, however, photons scattered out of the outward-directed beam are not fully replenished, leading to the observed cyclotron absorption feature. This process is analogous to the formation of absorption lines in the solar spectrum \citep{1979PhRvD..19.1684V}.

Assuming that the accreting gas is composed of pure, fully ionized hydrogen, we can use Equations (7) and (11) of \citet{1987ApJ...312..666A} to obtain the cyclotron photon emissivity in cgs units
\begin{align}\label{eq: cyc_emission_rate_1}
\dot{n}_{\epsilon}^{\rm cyc}
&=2.1\times10^{36}\, \rho^2 B_{12}^{-3/2}
H\left(\frac{\epsilon_{\rm c}}{k_B T_e}\right)
\nonumber \\
&\quad \times
\exp\left(-\frac{\epsilon_{\rm c}}{k_B T_e}\right)
\delta\left( \epsilon-\epsilon_{\rm c}\right) \,,
\end{align}
where $B_{12}(r)\equiv B(r)/(10^{12}\,{\rm G})$, with $B(r)$ given by Equation~(\ref{eq: dipole_B}), and
$\epsilon_{\rm c}(r)$ is the local cyclotron energy defined by Equation~(\ref{eq: cyclotron_energy}). The function $H(x)$, with $x\equiv \epsilon_{\rm{c}}/(k_{B} T_e) $, is defined by
\begin{equation}
H(x) \equiv
\begin{cases}
0.15 \, \sqrt{7.5} \,, & x \ge 7.5 
\,, \\
0.15 \, \sqrt{x} \,, & x < 7.5 \,.
\end{cases}
\end{equation}
The total cyclotron cooling rate is obtained by multiplying Equation (\ref{eq: cyc_emission_rate_1}) by the photon energy $\epsilon$ and integrating over all energies, which yields, in cgs units,
\begin{align}\label{eq: cyc_cooling_rate}
\dot{U}_{\rm cyc}^{\rm emit}
&=-2.1\times10^{36}\, \rho^2 B_{12}^{-3/2} \epsilon_{\rm c}
H\left(x\right)
\nonumber \\
&\quad \times
\exp\left( -\frac{\epsilon_{\rm c}}{k_B T_e}\right) \,,
\end{align}
where the negative sign indicates this is a cooling process for the electrons.

\subsubsection{Compton Heating and Cooling}
\label{sec: Compton_I}

Compton scattering plays a fundamental role in the formation of the emergent X-ray spectrum. It is critically important in establishing the radial variation of the electron temperature profile through the exchange of energy between photons and electrons. Equation (7.36) from \citet{1979rpa..book.....R} gives the mean change in the photon energy $\epsilon$ during a single scattering as
\begin{equation}\label{eq: Comp_emission_rate_1}
\left<\Delta\epsilon\right> = \frac{\epsilon}{m_e c^2}
\left(4 k_{B} T_e - \epsilon \right) \,,
\end{equation}
and the associated mean rate of change of the photon energy is therefore
\begin{equation}\label{eq: Comp_emission_rate_2}
\left.
\left\langle
\frac{{\rm d}\epsilon}{{\rm d}t}
\right\rangle
\right|_{\rm Comp}
= n_e \bar{\sigma} c
\left\langle \Delta \epsilon \right\rangle \,,
\vspace{0.3em}
\end{equation}
where $(n_e \bar{\sigma} c)^{-1}$ denotes the mean-free time between scatterings for the photons, and $\bar{\sigma}$ is the angle-averaged electron scattering cross section \citep{2007ApJ...654..435B, 2022ApJ...939...67B}. The corresponding rate of change of the electron energy density due to Compton scattering can therefore be written as
\begin{equation}\label{eq: Comp_emission_rate_3}
\dot U_{\rm Comp} = - n_e \bar\sigma c \int_0^\infty \epsilon^2 f(r,\epsilon)
\, \left<\Delta\epsilon\right> \, d\epsilon \,,
\end{equation}
where the distribution function, $f(r,\epsilon)$, is the solution to the photon transport equation introduced in \citet{2017ApJ...835..129W}, which is related to the total radiation number density, $n_{\rm rad}$, and energy density, $U_{\rm rad}$, via
\begin{align}\label{eq: Comp_emission_rate_4}
n_{\rm rad}(r)
&=
\int_0^\infty \epsilon^2 f(r,\epsilon)\,{\rm d}\epsilon,
\nonumber \\
U_{\rm rad}(r)
&=
\int_0^\infty \epsilon^3 f(r,\epsilon)\,{\rm d}\epsilon \,.
\end{align}
Combining Equations (\ref{eq: Comp_emission_rate_1}) and (\ref{eq: Comp_emission_rate_3}), we find that the net Compton energy-exchange rate for the electrons is given by
\begin{align}
\dot{U}_{\rm Comp} &=
\frac{n_e \bar{\sigma} c}{m_e c^2}
\left[\int_0^\infty \epsilon^4 f(r,\epsilon)\,
{\rm d}\epsilon 
\right. \nonumber \\
&\left.
\quad 
- 4 k_B T_e \int_0^\infty \epsilon^3 f(r,\epsilon)\, {\rm d}\epsilon \right] \,,
\end{align}
which vanishes if the electron temperature, $T_e$, is equal to the inverse-Compton temperature, $T_{\rm IC}$, defined by
\begin{equation}
T_{\rm IC}(r) \equiv \frac{1}{4k_{B}}\frac{\int_0^\infty
\epsilon^4 f(r,\epsilon) \, d\epsilon}{\int_0^\infty \epsilon^3 f(r,\epsilon)
\, d\epsilon} \,.
\end{equation}

For completeness, we summarize the Compton energy-exchange term used in the full radiation-hydrodynamical framework. In the present dynamical calculations, however, the photon distribution and the corresponding inverse-Compton temperature are not solved self-consistently, and the net Compton heating/cooling term is set to zero, as described in Section \ref{sec: solv_equations}. The Compton energy-exchange rate can be written as
\begin{equation}\label{eq: Comp_emission_rate_II}
\dot{U}_{\rm Comp}
=
n_e \bar{\sigma} c
\frac{4 k_B T_e}{m_e c^2}
\left[g(r)-1\right]
U_{\rm rad} \,,
\end{equation}
where we have introduced the temperature-ratio function
\begin{equation}\label{eq: Comp_g}
g(r) \equiv \frac{T_{\rm IC}}{T_e} \,.
\end{equation}
The sign of $\dot{U}_{\rm Comp}$ is determined by $g(r)$.  For $g(r)<1$, corresponding to $T_{\rm IC}<T_e$, the electrons lose energy through Compton cooling.  For $g(r)>1$, the electrons gain energy through inverse Compton heating.  Combining Equations (\ref{eq: eos_gas_pressure})--(\ref{eq: sound_speeds}) with Equation (\ref{eq: Comp_emission_rate_II}), we can express the Compton energy-exchange rate in terms of the mass density $\rho$, the electron sound speed $a_e$, and the radiation sound speed $a_{\rm rad}$ as
\begin{equation}\label{eq: Comp_emission_rate_III}
\dot{U}_{\rm Comp} =
\frac{ 4 \bar{\sigma} \left[g(r)-1\right] }{
m_e\, c\, \gamma_{e}\gamma_{\rm rad}
\left(\gamma_{\rm rad}-1\right) }
\rho^2 a_{\rm rad}^2 a_{e}^2 \,.
\end{equation}

\subsubsection{Electron--Ion Energy Exchange}
\label{sec: Electron-Ion}

Electrons and ions exchange thermal energy through Coulomb collisions. For the pure, fully ionized hydrogen plasma considered here, the ions are protons and hence their mass is $m_p$. Following \citet{2017ApJ...835..129W}, the net electron heating rate per unit volume due to electron--ion Coulomb coupling is
\begin{equation}\label{eq: electron_ion_exchange}
\begin{split}
\dot U_{ei} =
&\frac{3}{2}
\left(\frac{2}{\pi}\right)^{1/2}
\sigma_{\rm T} c^3 m_e n_e^2
\left(\frac{m_e}{m_p}\right)
\left(\frac{T_i-T_e}{T_{\rm eff}}\right)
\\
&\times
\left(
\frac{m_e c^2}{k_{\rm B}T_{\rm eff}}
\right)^{1/2}
\ln\Lambda_{\rm Coul}\,,
\end{split}
\end{equation}
where
\begin{equation}
T_{\rm eff}
=
T_e+\frac{m_e}{m_p}T_i \,.
\end{equation}
The Coulomb logarithm, $\ln\Lambda_{\rm Coul}$, is evaluated following \citet{2017ApJ...835..129W}. The sign of $\dot U_{ei}$ is determined by the temperature difference between the two species: energy is transferred from the ions to the electrons when $T_i>T_e$, whereas the transfer is reversed when $T_e>T_i$. The corresponding ion energy-exchange rate is $\dot U_i=-\dot U_{ei}$. Because the electron--ion equilibration timescale is short compared with the other characteristic energy-transport timescales, the two species remain close to thermal equilibrium throughout most of the accretion column. In the limiting case $T_e=T_i$, Equation~(\ref{eq: electron_ion_exchange}) gives $\dot U_{ei}=0$, as expected.

\section{Boundary Conditions and Thermal-Mound Definition}
\label{sec: boundary_conditions}

The thermal-mound surface is determined from the condition that the parallel free--free absorption optical depth reaches unity. The upper boundary conditions, described in Section \ref{sec: boundary_conditions_1}, involve the flow velocity and acceleration, the free-streaming radiation field, and conservation of the bulk-fluid momentum flux. The surface boundary conditions, described in Section \ref{sec: boundary_conditions_2}, account for flow stagnation and the attenuation of the total energy transport rate into the star. Near the column base, the decelerating flow forms a dense thermal mound that produces blackbody seed photons. Its upper surface, defined by $\tau_\parallel^{\rm ff}=1$, defines the thermalization surface discussed in Section \ref{sec: boundary_conditions_3}.

\subsection{Boundary Conditions at the Top of the Accretion Column}
\label{sec: boundary_conditions_1}

At the top of the accretion column, we assume that the inflowing gas is in free fall. The bulk velocity is therefore given by
\begin{equation}\label{eq: top_velocity}
v_{\rm top}= - \left(\frac{2GM_*}{r_{\rm top}}\right)^{1/2} \,,
\end{equation}
where the negative sign indicates that the flow is directed inward, opposite to the outward radial coordinate. The corresponding radial velocity gradient at the column top is
\begin{equation}\label{eq: top_velocity_gradient}
\left.
\frac{{\rm d}v}{{\rm d}r}
\right|_{r=r_{\rm top}} = \left( \frac{GM_*}{2r_{\rm top}^3} \right)^{1/2} \,.
\end{equation}

Following \citet{2017ApJ...835..129W}, we identify the upper boundary of the accretion column with the last-scattering surface for photons escaping through the top of the column. In analogy with stellar-atmosphere theory, this condition requires the parallel scattering optical depth from $r_{\rm top}$ to $r_c$ to be unity. Here $r_c$ is associated with the magnetic-equatorial plane and satisfies $r_c>r_{\rm top}$. Defining the parallel scattering optical depth, $\tau_{\parallel}^{\rm scat}$, so that it increases in the downward direction for bulk fluid entering at the top of the column and flowing downward, from $\tau_{\parallel}^{\rm scat} = 0$ at $r = r_{\rm top}$, we have
\begin{equation}\label{eq: parallel_scatt_1}
\tau_{\parallel}^{\rm scat}(r)
=
\int_{r}^{r_{\rm top}} n_e(r')\sigma_\parallel\,{\rm d}r' \,,
\qquad R_*\le r \le r_{\rm top} \,.
\end{equation}
Since the top of the accretion column is the last scattering surface, we can also write
\begin{equation}\label{eq: parallel_scatt_2}
\int^{r_{c}}_{r_{\rm top}} n_e(r') \sigma_{\parallel}\,{\rm d}r'  = 1 \,,
\end{equation}
where $r_c > r_{\rm top}$.

Assuming free fall above the column top and using mass continuity in the funnel geometry, the optical-depth condition gives
 \begin{equation}\label{eq: top_tau_integrated}
\frac{2\sigma_\parallel \dot M}
{3m_p(2GM_*)^{1/2}}
\frac{R_*}{\Omega_*}
\left(r_{\rm top}^{-3/2}-r_c^{-3/2}\right)
=1 \,.
\end{equation}
Solving for the dimensional column-top radius, we obtain
\begin{equation}\label{eq: top_radius}
r_{\rm top}
=
\left[
r_c^{-3/2} + \frac{3m_p(2GM_*)^{1/2}} {2\sigma_\parallel \dot M}
\frac{\Omega_*}{R_*}
\right]^{-2/3} \,,
\end{equation}
where
\begin{equation}\label{eq: critical_radius}
r_c= \frac{2R_*}{3\sin^2\theta_2} \,.
\end{equation}
This relation allows us to determine $r_{\rm top}$ self-consistently for a given set of model parameters, including $\Omega_*$, $\theta_2$, and $\sigma_{\parallel}$.

Because the electron--ion equilibration timescale is short compared with the other characteristic energy-transport timescales, we assume $T_{e,\rm top}=T_{i,\rm top}$ at the column top. The electron and ion energy equations are subsequently evolved independently throughout the column, with Coulomb coupling maintaining $T_e \simeq T_i$ over most of the flow. The relevant energy-transport timescales are discussed in detail in Section~7.5 of \citet{2017ApJ...835..129W}. The electron and ion sound speeds at the top of the column are given by
\begin{equation}\label{eq: top_gas_sound_speed}
a_{e, \rm top}^2 =\frac{\gamma_{e} 
k_B T_{e,\rm top}}{m_p} \,, 
\quad
a_{i, \rm top}^2 =\frac{\gamma_{i} 
k_B T_{i,\rm top}}{m_p} \,, 
\end{equation}
and therefore our assumption that $T_{e, \rm top} = T_{i, \rm top}$, leads to the relation
\begin{equation}\label{eq: top_gas_sound_speed_2}
a_{e, \rm top} = 
\left(\frac{\gamma_{e}}{\gamma_{i}} \right)^{1/2}
a_{i, \rm top} \,.
\end{equation}

The radial component of the radiation energy flux, averaged over the cross section of the accretion column at radius $r$, is given by \citep{1992ApJ...397...88B}
\begin{equation}\label{eq: rad_flux}
F_{\rm rad}(r)
=
-\frac{c}{3n_e\sigma_{\parallel}}
\frac{{\rm d}U_{\rm rad}}{{\rm d}r}
+\frac{4}{3}vU_{\rm rad} \,,
\end{equation}
where the first term on the right-hand side represents spatial diffusion and the second term represents advection with the accreting flow. At the top of the accretion column, the optical depth becomes sufficiently small that photon transport transitions from diffusion to free streaming. We therefore impose the free-streaming boundary condition at $r=r_{\rm top}$ by replacing the diffusive flux term in Equation (\ref{eq: rad_flux}) with the free-streaming flux,
\begin{equation}\label{eq: free_streaming}
-\frac{c}{3n_e\sigma_{\parallel}}
\frac{{\rm d}U_{\rm rad}}{{\rm d}r}
\rightarrow
c\,U_{\rm rad}\,,
\qquad r \rightarrow r_{\rm top} \,.
\end{equation}
Substituting this condition into Equation (\ref{eq: rad_flux}), we obtain the radiation energy flux at the upper surface of the column,
\begin{equation}\label{eq: rad_flux_top}
F_{\rm rad}(r_{\rm top})
=
\left(c+\frac{4}{3}v_{\rm top}\right)
U_{\rm rad}(r_{\rm top}) \,.
\end{equation}

We next evaluate the corresponding total energy transport rate. Using Equation (\ref{eq: energy_transport_rate}), together with the equation-of-state relations in Equation (\ref{eq: eos_energy_density}), the sound-speed definitions in Equation (\ref{eq: sound_speeds}), and the mass-continuity relation in Equation (\ref{eq: mass_continuity}), we obtain
\begin{equation}\label{eq: 77}
\begin{split}
\dot E(r)
= A(r)F(r)
= \dot{M} \bigg[
& \frac{F_{\rm rad}}{\rho |v|}
- \frac{v^2}{2}
- \frac{a_i^2}{\gamma_i-1}
\\
& - \frac{a_e^2}{\gamma_e-1}
+ \frac{GM_*}{r}
\bigg] \,.
\end{split}
\end{equation}

The upper boundary value of the total energy transport rate follows by applying the free-streaming condition in Equation (\ref{eq: free_streaming}) to the radiation flux and by using Equations (\ref{eq: eos_energy_density}) and (\ref{eq: sound_speeds}) to express $U_{\rm rad}$ in terms of the radiation sound speed $a_{\rm rad}$. Together with the upper boundary conditions for the bulk flow, this gives
\begin{equation}\label{eq: top_flux}
\begin{aligned}
\dot E_{\rm top}
& =
\left.\dot E\right|_{r=r_{\rm top}}
\\
&= -\dot M
\bigg[\left(\frac{\gamma_e}{\gamma_e-1}
+ \frac{\gamma_i}{\gamma_i-1}\right)
\frac{a_{i,{\rm top}}^2}{\gamma_i}
\\
&\qquad
+\left(\frac{c}{v_{\rm top}}+\frac{4}{3}\right)
\frac{a_{{\rm rad},{\rm top}}^2}
{\gamma_{\rm rad}(\gamma_{\rm rad}-1)} \bigg] \,.
\end{aligned}
\end{equation}

\subsection{Boundary Conditions at the Stellar Surface}
\label{sec: boundary_conditions_2}


At the stellar surface, we impose a vanishing net radial radiation flux, corresponding to the mirror boundary condition adopted for the base of the accretion column \citep{2022ApJ...939...67B}
\begin{equation}\label{eq: surface_rad_flux}
F_{\rm rad}(r)
=
-\frac{c}{3n_e\sigma_{\parallel}}
\frac{{\rm d}U_{\rm rad}}{{\rm d}r}
+\frac{4}{3}vU_{\rm rad}=0\,,
\qquad r \rightarrow R_* \,.
\end{equation}

Combining this condition with Equation \ref{eq: 77}, the total energy-transport rate at the stellar surface is
\begin{equation}\label{eq: surface_rad_flux_2}
\dot E_*
=
\dot M
\left[
-\frac{v_*^2}{2}
-\frac{a_{i,*}^2}{\gamma_i-1}
-\frac{a_{e,*}^2}{\gamma_e-1}
+\frac{GM_*}{R_*}
\right] \,,
\end{equation}
where the subscript $*$ denotes quantities evaluated at $r=R_*$. Equation~(\ref{eq: surface_rad_flux_2}) shows that, for a numerical solution retaining a finite residual velocity and finite gas enthalpy, the surface energy-transport rate is slightly smaller than the pure gravitational contribution $GM_*\dot M/R_*$. The latter is recovered only in the ideal cold-stagnation limit.

Formally, exact stagnation requires $v=0$ at the stellar surface. In practice, imposing $v=0$ leads to a divergent mass density through the continuity equation. We therefore require the flow to approach stagnation numerically,
\begin{equation}\label{eq: surface_speed}
\lim_{r\rightarrow R_*}|v(r)|\lesssim 0.02c \,.
\end{equation}
The lower-boundary solution is selected by requiring Equation~(\ref{eq: surface_rad_flux}) to be satisfied while maintaining the approximate stagnation condition in Equation~(\ref{eq: surface_speed}).


\subsection{Location of the Thermal Mound Surface}
\label{sec: boundary_conditions_3}

As the flow decelerates near the base of the accretion column, the density increases and free--free absorption becomes increasingly effective, leading to the formation of a dense thermal mound \citep{1973NPhS..246....1D}. This mound, with a characteristic temperature of $10^7$--$10^8\,{\rm K}$, provides the blackbody seed photons that are Comptonized throughout the column and contribute to the emergent spectrum. The upper surface of the thermal mound is located at $r=r_{\rm th}$, which is defined as the radius where the Rosseland mean free--free optical depth measured along the column reaches unity \citep{2007ApJ...654..435B, 2022ApJ...939...67B}.

The parallel Rosseland mean free--free optical depth is computed as
\begin{equation}\label{eq: taurosseland}
\tau_\parallel^{\rm ff}(r)
=
\int_r^{r_{\rm top}} \alpha_{\rm R}(r')\,{\rm d}r' \,,
\end{equation}
where $\alpha_{\rm R}$ is the Rosseland mean free--free absorption coefficient for fully ionized hydrogen. This definition gives
\begin{equation}
\tau_\parallel^{\rm ff}(r_{\rm top}) = 0 \,,
\end{equation}
at the top of the column. The thermal mound surface is then determined by the condition
\begin{equation}\label{eq: absorption_unity}
\tau_\parallel^{\rm ff}(r_{\rm th})
= \int_{r_{\rm th}}^{r_{\rm top}}
\alpha_{\rm R}(r')\,{\rm d}r' = 1 \,.
\end{equation}
Below this surface, $\tau_\parallel^{\rm ff}>1$, so free--free emission and absorption are approximately balanced. This balance is not exact, however, because photons can still escape laterally through the column walls.

\section{Solving the Coupled System Equations}
\label{sec: solv_equations}

In this section, we describe the dimensionless formulation and the numerical procedure used to solve the five coupled first-order equations. For each source, the solution is obtained by integrating from the column top to the stellar surface and selecting physically acceptable solutions that approximately satisfy the lower boundary conditions at the stellar surface.

It is convenient to work in terms of nondimensional radius, flow velocity, sound speed, and total energy transport rate variables by introducing the quantities
\begin{equation}\label{eq: convertingr}
\begin{aligned}
\tilde r &= \frac{r}{R_g}\,,
\qquad
\tilde v = \frac{v}{c}\,,
\qquad
\tilde a_e = \frac{a_e}{c}\,,
\\
\tilde a_i &= \frac{a_i}{c}\,,
\qquad
\tilde a_{\rm rad} = \frac{a_{\rm rad}}{c}\,,
\qquad
\tilde{\mathscr{E}} = \frac{\dot E}{\dot M c^2} \,,
\end{aligned}
\end{equation}
where $R_g$ is the gravitational radius, defined by
\begin{equation}\label{eq: R_g}
R_g \equiv \frac{G M_*}{c^2} \,.
\end{equation}
The computational domain extends from the top of the accretion column, at radius $\tilde r = \tilde r_{\rm top}$, down to the stellar surface at dimensionless radius $\tilde r_{\rm sur}=4.836$, which is based on canonical values for the stellar mass, $M_{*}=1.4\, M_{\odot}$, and the stellar radius, $R_{*}=10\,\rm km$.

The coupled set of conservation equations corresponding to Equations~(\ref{eq: rad_speed}), (\ref{eq: electron_speed}), (\ref{eq: ion_speed}), (\ref{eq: bulk_velocity}), and (\ref{eq: total_energy}) are given by
\begin{subequations}\label{eq: dimensionless_odes}
\begin{align}
\frac{{\rm d} {\tilde a_{\rm rad}}}{{\rm d}{\tilde r}}
&= \frac{\tilde a_{\rm rad}}{2}
\left(\frac{3}{\tilde r} + \frac{1}{\tilde v}
\frac{{\rm d}{\tilde v}}{{\rm d}{\tilde r}}\right)
\nonumber \\
&\quad
- \frac{\sigma_\parallel R_g}{2m_p c}
\frac{\dot M}{A} \frac{\gamma_{\rm rad}}{\tilde a_{\rm rad}}
\left(\tilde{\mathscr{E}} + \frac{\tilde v^2}{2}
+ \frac{\tilde a_e^2}{\gamma_e-1}
\right.
\nonumber \\
&\quad\left.
+ \frac{\tilde a_i^2}{\gamma_i-1}
+ \frac{\tilde a_{\rm rad}^2}{\gamma_{\rm rad}-1}
- \frac{1}{\tilde r}
\right) \,,
\label{eq: RadiationODEtilde}
\\[6pt]
\frac{{\rm d} {\tilde a_i}}{{\rm d}{\tilde r}}
&= \frac{(1-\gamma_i)\tilde a_i}{2}
\left[ \frac{3}{\tilde r}
+ \frac{1}{\tilde v}
\frac{{\rm d}{\tilde v}}{{\rm d}{\tilde r}}
+ \frac{R_g}{c^2} \frac{A}{\dot M}
\frac{\gamma_i \dot U_i}{\tilde a_i^2}
\right] \,,
\label{eq: IonODEtilde}
\\[6pt]
\frac{{\rm d}{\tilde a_e}}{{\rm d}{\tilde r}}
&= \frac{(1-\gamma_e)\tilde a_e}{2}
\left[ \frac{3}{\tilde r}
+ \frac{1}{\tilde v} \frac{{\rm d}{\tilde v}}{{\rm d}{\tilde r}}
+ \frac{R_g}{c^2} \frac{A}{\dot M}
\frac{\gamma_e \dot U_e}{\tilde a_e^2}
\right] \,,
\label{eq: ElectronODEtilde}
\\[6pt]
\frac{{\rm d} \tilde{\mathscr{E}}}{{\rm d} \tilde r}
&=
\frac{\tilde a_{\rm rad}^2}
{c(\ell_2-\ell_1)\gamma_{\rm rad}(\gamma_{\rm rad}-1)\tilde v}
\nonumber \\
&\quad \times
\left( \frac{R_*^3}{R_g\,\tilde r^3}
\right)^{1/2}
\min\left(c,\frac{c}{\tau_\perp}\right),
\label{eq: EnergyODEtilde}
\\[6pt]
\frac{{\rm d} {\tilde v}}{{\rm d} {\tilde r}}
&= \frac{\tilde v}
{\tilde v^2-\tilde a_i^2-\tilde a_e^2}
\bigg\{ \frac{3(\tilde a_i^2+\tilde a_e^2)}{\tilde r}
- \frac{1}{\tilde r^2}
\nonumber \\
&\quad
+ \frac{\sigma_\parallel R_g}{m_p c}
\frac{\dot M}{A}
\left( \tilde{\mathscr{E}}
+ \frac{\tilde v^2}{2}
+ \frac{\tilde a_i^2}{\gamma_i-1}
\right.
\nonumber \\
&\quad \left.
+ \frac{\tilde a_e^2}{\gamma_e-1}
+ \frac{\tilde a_{\rm rad}^2}{\gamma_{\rm rad} - 1}
- \frac{1}{\tilde r}
\right)
\nonumber \\
&\quad
+ \frac{R_g}{c^2} \frac{A}{\dot M}
\left[(\gamma_i-1)\dot{U}_i + (\gamma_e-1)\dot{U}_e \right]
\bigg\} \,.
\label{eq: VelocityODEtilde}
\end{align}
\end{subequations}

In solving the coupled hydrodynamical equations, we adopt the simplifying approximation $g(\tilde r)=1$ in the Compton energy-exchange term (see Equation~\ref{eq: Comp_g}). Since $g(\tilde r)=T_{\rm IC}/T_e$, this corresponds to setting $T_{\rm IC}(\tilde r)=T_e(\tilde r)$ throughout the accretion column, so that the net Compton heating and cooling term vanishes. Consequently, the angle-averaged scattering cross section $\bar{\sigma}$ does not affect the present hydrodynamical solutions. We retain $\bar{\sigma}$ as a model parameter for consistency with the full radiation-transport framework, in which it enters the Compton energy exchange and photon transport. A self-consistent calculation of a spatially varying $T_{\rm IC}$ will be addressed in future work by coupling the hydrodynamical solution to the photon-transport calculation.  This approximation differs from the fully coupled radiation-hydrodynamical and photon-transport treatment of \citet{2017ApJ...835..129W, 2017ApJ...835..130W}, in which the inverse-Compton temperature is determined iteratively from the photon distribution. The present calculations therefore focus on the hydrodynamical structure under a simplified Compton energy-exchange prescription, rather than on the self-consistent formation of the emergent X-ray spectrum.

Using the dimensionless variables defined in Equations (\ref{eq: convertingr}) and (\ref{eq: R_g}), the free-fall velocity at the top of the column becomes
\begin{equation}\label{eq: vtop_tilde}
\tilde v_{\rm top} = -\left(\frac{2}{\tilde r_{\rm top}}\right)^{1/2} \,,
\end{equation}
and the corresponding local acceleration is
\begin{equation}\label{eq: dvdr_top_tilde}
\left.
\frac{{\rm d}\tilde v}{{\rm d}\tilde r}
\right|_{\tilde r=\tilde r_{\rm top}}
= -\left. \frac{1}{\tilde r^2 \tilde v} \right|_{\tilde r=\tilde r_{\rm top}} \,.
\end{equation}

The lower spatial boundary condition is imposed at the stellar surface. We require the net photon energy flux to vanish there (see Equation \ref{eq: surface_rad_flux}), so that no net radiative energy is transported into or out of the star at the base of the column \citep{2022ApJ...939...67B}. This condition ensures that the emergent X-ray luminosity is powered by accretion within the column, rather than by an additional energy source from the stellar interior. We therefore relate the mass accretion rate to the X-ray luminosity using
\begin{equation}\label{eq: mdot_lx}
\dot M =
\frac{L_{\rm X} R_*}{GM_*} \,.
\end{equation}
The constrained source parameters used in this work are listed in Table~\ref{tab: constrained_parameters}.

To close the system of five coupled conservation equations, we also specify the incident electron, ion, and radiation Mach numbers at the top of the accretion column. We first impose the radiation boundary condition through the incident radiation Mach number. The incident radiation Mach number at the column top is defined as
\begin{equation}\label{eq: rad_Mach}
\mathcal{M}_{\rm rad,top}
= \frac{|\tilde v_{\rm top}|}{\tilde a_{\rm rad,top} } \,.
\end{equation}
In the numerical calculation, $M_{\rm rad,top}$ serves as a shooting parameter. For each source, its value is adjusted within a physically acceptable range so that the inward integration satisfies the mirror boundary condition $F_{\rm rad}(R_*)=0$ at the stellar surface, while the residual surface velocity remains consistent with the approximate stagnation requirement in Equation~(\ref{eq: surface_speed}).

The gas boundary condition is specified analogously. However, the incident ion Mach number, $M_{i,\rm top}$, is not treated as an independent free parameter. Following the Appendix of \citet{2017ApJ...835..129W}, it is determined by enforcing momentum conservation and matching the hydrodynamical velocity gradient to the free-fall gradient at the column top.



The gas boundary condition is specified analogously. However, the
incident ion Mach number, $\mathcal{M}_{i,{\rm top}}$, is not treated as an independent free parameter. Following the Appendix of \citet{2017ApJ...835..129W}, it is determined by enforcing momentum conservation and matching the hydrodynamical velocity gradient to the free-fall gradient at the column top. The resulting implicit condition can be written as
\begin{equation}\label{eq:ion_mach_condition}
\begin{split}
\frac{1}{\mathcal{M}_{i,{\rm top}}^2}
=&
-\frac{27}{56\mathcal{M}_{\rm rad,top}^2}
\left[
1-
\left(
\frac{\tilde r_{\rm top}}{\tilde r_c}
\right)^{3/2}
\right]^{-1}
\\
&-
\frac{1}{7}
\frac{R_g^4\Omega_*}
{\dot M c^2 R_*}
\tilde r_{\rm top}^{5}
\dot U_{e,{\rm top}} \,,
\end{split}
\end{equation}
where $\dot U_{e,{\rm top}}$ is the net electron energy-exchange rate evaluated at the column top. Since $\dot U_{e,{\rm top}}$ depends on the electron temperature and hence on $\mathcal{M}_{i,{\rm top}}$, Equation~(\ref{eq:ion_mach_condition}) is solved numerically using a root finder. For the adopted models, we obtain
$\mathcal{M}_{i,{\rm top}}=218.81$, $137.92$, and $79.60$ for NGC 300 ULX-1, Swift J0243.6$+$6124, and SMC X-3, respectively.

The ion sound speed at the top of the column is then
\begin{equation}\label{eq: ion_Mach}
\tilde a_{i,{\rm top}}
=
\frac{|\tilde v_{\rm top}|}
{\mathcal{M}_{i,{\rm top}}}\,.
\end{equation}
The electron sound speed follows from the condition $T_{e,{\rm top}}=T_{i,{\rm top}}$,
\begin{equation}\label{eq: electron_Mach}
\tilde a_{e,{\rm top}}
=
\left(
\frac{\gamma_e}{\gamma_i}
\right)^{1/2}
\tilde a_{i,{\rm top}}\,.
\end{equation}

The final upper boundary condition is specified by the total energy transport rate. Converting Equation (\ref{eq: top_flux}) into dimensionless form gives
\begin{equation}\label{eq: Etildetop}
\begin{aligned}
\tilde{\mathscr{E}}_{\rm top}
=&\
\left( \frac{\gamma_e}{1-\gamma_e}
+ \frac{\gamma_i}{1-\gamma_i} \right)
\frac{\tilde a_{i,{\rm top}}^2}{\gamma_i}
\\
&+
\left( \frac{1}{\tilde v_{\rm top}} +
\frac{4}{3} \right) \frac{\tilde a_{{\rm rad},{\rm top}}^2}
{\gamma_{\rm rad}(1-\gamma_{\rm rad})} \,.
\end{aligned}
\end{equation}
With these upper and lower boundary conditions specified, the coupled conservation equations can be solved for each ULXP once the corresponding source-dependent parameters are chosen. The adopted model parameters and constrained source parameters are summarized in Tables \ref{tab: model_parameters} and \ref{tab: constrained_parameters}, respectively.

\section{Astrophysical Applications}
\label{sec: astro_apply}

In this section, we apply the self-consistent radiation-hydrodynamical model to three super-Eddington X-ray pulsars: NGC 300 ULX-1, Swift J0243.6$+$6124, and SMC X-3. The quantities listed in Table~\ref{tab: model_parameters} are adopted
model parameters. Among them, $\sigma_\parallel$, the column geometry, $\mathcal{M}_{\rm rad,top}$, and $B_*$ enter the present dynamical calculation directly. Among these quantities, $M_{\rm rad,top}$ is determined through the shooting procedure described in Section \ref{sec: solv_equations} so that the lower radiation-flux boundary condition is satisfied. The angle-averaged cross section $\bar{\sigma}$ is retained for consistency with the full radiation-transport framework but does not affect the current hydrodynamical solutions because the net Compton energy-exchange term is set to zero. Since the present work does not perform spectral fitting, these parameters should not be interpreted as unique best-fit values. Instead, they are chosen to obtain physically acceptable hydrodynamical solutions that satisfy the imposed boundary conditions and remain consistent with the available observational constraints.

The constrained source parameters are listed in Table \ref{tab: constrained_parameters}. These include the stellar mass and radius, distance, X-ray luminosity, corresponding mass accretion rate, and perpendicular scattering cross section. The mass accretion rate is estimated from the observed luminosity using Equation (\ref{eq: mdot_lx}).

The opening angles determine the column solid angle and therefore the density scale. The parallel scattering cross section controls the radiative force along the column, while $\mathcal{M}_{\rm rad,top}$ sets the radiation pressure at the upper boundary. The adopted magnetic field strength is constrained independently for each source. In this paper, we focus on how these parameters determine the self-consistent dynamical structure of the accretion column.

\begin{deluxetable}{lccc}
\tablewidth{0pt}
\tablecaption{Adopted model parameters for NGC 300 ULX-1, Swift J0243.6$+$6124, and SMC X-3. The angle-averaged scattering cross section $\bar{\sigma}$ is retained for consistency with the full radiation-transport framework. It does not affect the present hydrodynamical solutions because the net Compton energy-exchange term is set to zero.
\label{tab: model_parameters}
}
\tablehead{
\colhead{Parameter} &
\colhead{NGC 300 ULX-1} &
\colhead{Swift J0243.6$+$6124} &
\colhead{SMC X-3}
}
\startdata
\shortstack[l]{\rule{0pt}{2.6ex}Angle-averaged scattering\\[1pt]
cross section, $\bar{\sigma}/\sigma_{\rm T}$}
& $9.50\times10^{-4}$
& $5.50\times10^{-4}$
& $4.00\times10^{-4}$ \\[2pt]
\tableline
\shortstack[l]{\rule{0pt}{2.6ex}Parallel scattering\\[1pt]
cross section, $\sigma_{\parallel}/\sigma_{\rm T}$}
& $1.20\times10^{-4}$
& $1.37\times10^{-4}$
& $2.11\times10^{-4}$ \\[2pt]
\tableline
Inner polar cap radius, $\ell_{1}\,({\rm m})$
& $500$
& $550$
& $500$ \\[2pt]
\tableline
Outer polar cap radius, $\ell_{2}\,({\rm m})$
& $900$
& $750$
& $700$ \\[2pt]
\tableline
\shortstack[l]{\rule{0pt}{2.6ex}Incident radiation Mach\\[1pt]
number, $\mathcal{M}_{\rm rad,top}$}
& $1.38$
& $1.45$
& $1.50$ \\[2pt]
\tableline
Surface magnetic field, $B_*$ ($10^{13}\,{\rm G}$)
& $0.6$
& $1.6$
& $4.0$ \\
\enddata
\end{deluxetable}

\begin{deluxetable}{lcccc}
\tablewidth{0pt}
\tablecaption{
Constrained parameters adopted for NGC 300 ULX-1, Swift J0243.6$+$6124, and SMC X-3.  For Swift J0243.6$+$6124, we retain the earlier luminosity estimate of $L_{\rm X}=2.0\times10^{39}\,{\rm erg\,s^{-1}}$ at $D=7\,{\rm kpc}$ as the input for our fiducial model. The updated distance and the corresponding luminosity estimate are provided in Table~\ref{tab: Table_1}.
\label{tab: constrained_parameters}
}
\tablehead{
\colhead{Parameter} &
\colhead{NGC 300 ULX-1} &
\colhead{Swift J0243.6$+$6124} &
\colhead{SMC X-3} &
\colhead{Unit}
}
\startdata
$R_*$
& $10$
& $10$
& $10$
& ${\rm km}$ \\[3pt]
\tableline
$M_*$
& $1.4$
& $1.4$
& $1.4$
& $M_\odot$ \\[3pt]
\tableline
$D$
& $2.0$
& $0.007$
& $0.062$
& ${\rm Mpc}$ \\[3pt]
\tableline
$L_{\rm X}$
& $4.7\times10^{39}$
& $2.0\times10^{39}$
& $1.2\times10^{39}$
& ${\rm erg\,s^{-1}}$ \\[3pt]
\tableline
$\dot{M}$
& $2.52\times10^{19}$
& $1.07\times10^{19}$
& $6.54\times10^{18}$
& ${\rm g\,s^{-1}}$ \\[3pt]
\tableline
$\sigma_{\perp}$
& $\sigma_{\rm T}$
& $\sigma_{\rm T}$
& $\sigma_{\rm T}$
& ${\rm cm^2}$ \\
\enddata
\end{deluxetable}

\subsection{NGC 300 ULX-1}
\label{sec: case_1}

\begin{figure*}[ht]
    \centering
    \includegraphics[width=15cm]{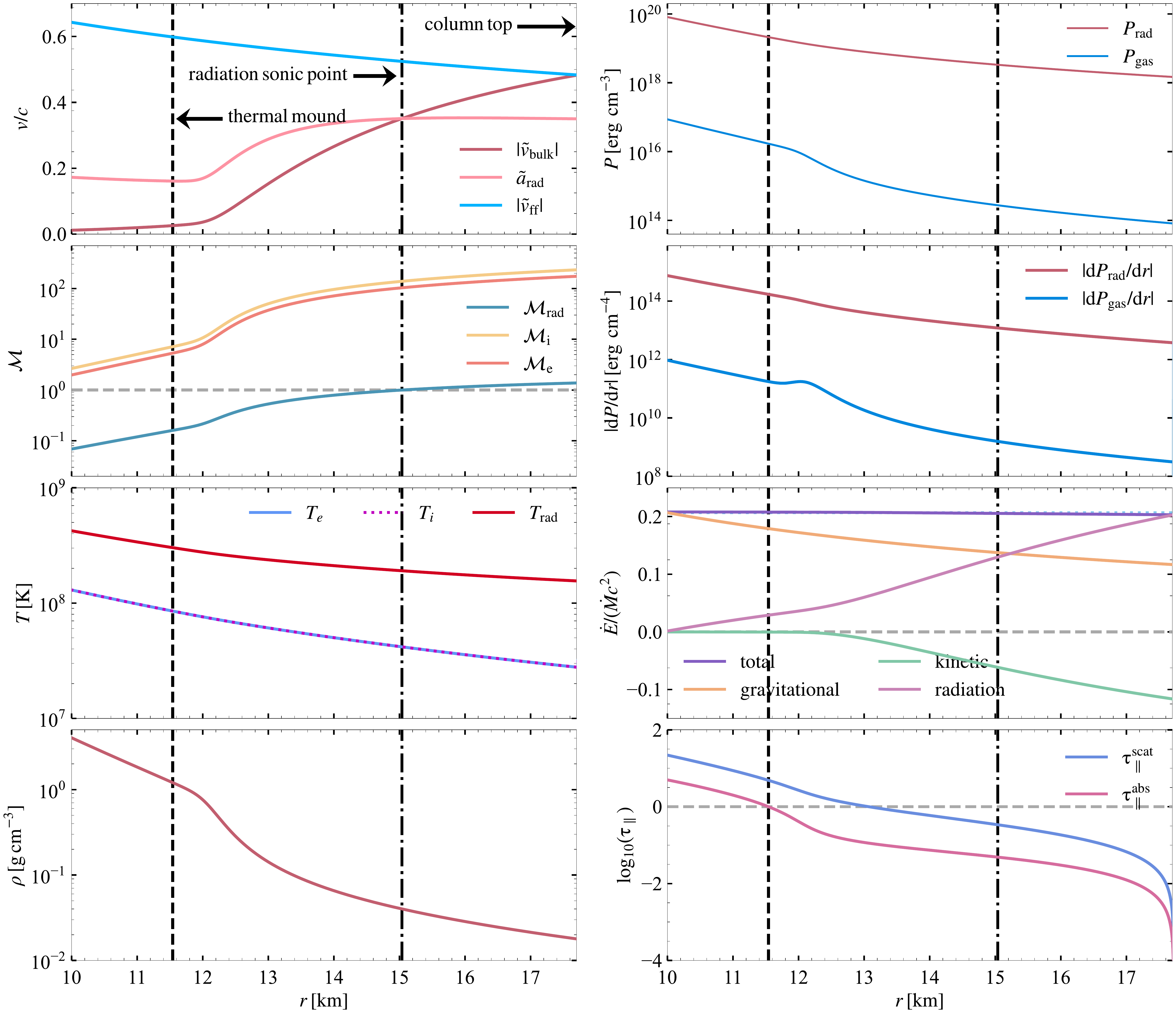}
    \caption{ Profiles of physical quantities within the accretion column of NGC 300 ULX-1. The first column shows the bulk speed, Mach numbers, temperatures, and mass density. The electron and ion energy equations are evolved separately in our two-fluid formulation. Their temperature profiles are nearly indistinguishable because rapid Coulomb energy exchange maintains the electrons and ions close to thermal equilibrium throughout most of the column. The radiation temperature, $T_{\rm rad}$, is the equivalent temperature associated with the local radiation energy density, defined by $U_{\rm rad}=a_{\rm R}T_{\rm rad}^{4}$, where $a_{\rm R}$ is the radiation constant. It is introduced only as a convenient diagnostic of the radiation energy density and does not imply that the local radiation field is Planckian. The second column shows the gas and radiation pressures, their radial gradients, the energy-transport components, and the parallel scattering and free--free absorption optical depths. In all panels, the vertical dashed line marks the thermal-mound surface, defined by $\tau_{\parallel}^{\rm ff}=1$, while the vertical dash-dotted line marks the radiation sonic surface, where $a_{\rm rad}=|v|$.}
    \label{fig: NGC300}
\end{figure*}

Figure~\ref{fig: NGC300} shows the calculated accretion-column structure
of NGC 300 ULX-1. We adopt an X-ray luminosity of $L_{\rm X}=4.7\times10^{39}\,{\rm erg\,s^{-1}}$ \citep{Carpano:2018pot,Vasilopoulos:2018vqh,Walton:2018ecy}, corresponding to a mass accretion rate of $\dot M=2.52\times10^{19}\,{\rm g\,s^{-1}}$ for the adopted stellar mass and radius. The adopted model parameters are $\bar{\sigma}/\sigma_{\rm T}=9.50\times10^{-4}$, $\sigma_{\parallel}/\sigma_{\rm T}=1.20\times10^{-4}$, $\ell_1=500\,{\rm m}$, $\ell_2=900\,{\rm m}$, $\mathcal{M}_{\rm rad,top}=1.38$, and $B_*=6\times10^{12}\,{\rm G}$. The resulting column top is located at $r_{\rm top}=17.7\,{\rm km}$, corresponding to a column height of
$7.7\,{\rm km}$ above the stellar surface. In Figure~\ref{fig: NGC300}, the column top is located on the right-hand side of each panel, whereas the stellar surface is located on the left-hand side.

Although the electron and ion energy equations are retained separately in our two-fluid formulation, Coulomb collisions efficiently exchange
energy between the two species. The electron--ion equilibration timescale is substantially shorter than the other characteristic energy-transport timescales in luminous X-ray pulsar accretion columns
\citep{2017ApJ...835..129W}. Consequently, the numerical solutions satisfy $T_e\simeq T_i$ throughout the column, and the two temperature profiles are nearly indistinguishable in Figure~\ref{fig: NGC300}. We emphasize that the equality of the two temperatures is not imposed as a single-temperature closure; rather, it results from the strong electron--ion coupling in the separately evolved energy equations.

For comparison with the gas temperatures, we also show an equivalent radiation temperature, $T_{\rm rad}$, defined from the local radiation energy density according to
\begin{equation}
T_{\rm rad}
\equiv
\left(\frac{U_{\rm rad}}{a_{\rm R}}\right)^{1/4}
=
\left[
\frac{P_{\rm rad}}
{(\gamma_{\rm rad}-1)a_{\rm R}}
\right]^{1/4} \,,
\end{equation}
where $a_{\rm R}$ is the radiation constant. This quantity provides a
convenient measure of the local radiation energy density and does not
imply that the radiation field is locally Planckian. In particular,
$T_{\rm rad}$ is not used as a thermodynamic closure relation and should be distinguished from the inverse-Compton temperature, $T_{\rm IC}$, which depends on the spectral shape of the radiation field. A
self-consistent calculation of $T_{\rm IC}(r)$ requires iteration between the hydrodynamical and photon-transport solutions. Since such an iteration is not performed in the present dynamical calculation, the net Compton heating/cooling term is omitted here.

The radiation sonic surface, defined by $a_{\rm rad}=|v|$, is located at $r_{\rm sonic}=15.032\,{\rm km}$. Below this radius, the flow becomes subsonic with respect to radiation-pressure perturbations and enters an extended sinking regime \citep{1976MNRAS.175..395B,1998ApJ...498..790B}. Radiation pressure provides the dominant dynamical support in this region and gradually decelerates the accreting material as it approaches the stellar surface. Mass continuity then requires the density to increase as the inflow slows, while the gas and radiation pressures also rise toward the base of the column.

The thermal-mound surface is determined from the condition $\tau_{\parallel}^{\rm ff}=1$ in Equation~(\ref{eq: absorption_unity}). For the adopted model parameters, we obtain $r_{\rm th}=11.534\,{\rm km}$, corresponding to a vertical extent
$h_{\rm th}\equiv r_{\rm th}-R_*=1.534\,{\rm km}$. The electron temperature at the thermal-mound surface is $T_e(r_{\rm th})=8.53\times10^{7}\,{\rm K}$. Below this radius, the free--free absorption optical depth increases rapidly, and the material becomes increasingly effective at absorbing and reprocessing the
radiation field. Defining $v_*\equiv v(R_*)$ as the bulk flow velocity at the stellar surface, the numerical solution gives $|v_*|=0.0119\,c$, indicating approximate stagnation at the lower computational boundary.

The Mach-number profiles show that the flow remains supersonic with respect to the electron and ion sound speeds at the lower boundary of the computational domain. A terminal gas-mediated discontinuous shock may therefore occur as the residual inflow merges with the stellar crust. However, the kinetic-energy contribution at the stellar surface is small: $|v_*|^2/2\simeq7\times10^{-5}$ in units of $c^2$, which is negligible compared with the radiation-dominated energy transport. We therefore do not model the detailed structure of this terminal shock, following the treatment adopted for Her X-1, Cen X-3, and LMC X-4 by
\citet{2017ApJ...835..129W}.

The energy-transport components shown in Figure~\ref{fig: NGC300} are normalized by $\dot M c^2$. For reference, the magnitude of the gravitational contribution is
\begin{equation}\label{eq: gravitational_energy_transport}
\frac{\dot E_g}{\dot M c^2}
=
\frac{GM_*\dot M/r}{\dot M c^2}
=
\frac{1}{\tilde r}\,.
\end{equation}
For the adopted stellar parameters, the stellar surface corresponds to
$\tilde r_{\rm sur}\simeq4.836$, giving $1/\tilde r_{\rm sur}\simeq0.2068$. Near the column top, the free-fall kinetic and gravitational terms nearly cancel, and the net energy transport is dominated by the radiation component. As the flow enters the sinking region and decelerates, the kinetic contribution becomes progressively smaller. Defining $\dot E_* \equiv \dot E(R_*)$ as the total energy-transport rate at the stellar surface, the numerical solution gives
\begin{equation}
\frac{\dot E_*}{\dot M c^2}=0.2066\,.
\end{equation}
This value is slightly smaller than the pure gravitational contribution, $1/\tilde r_{\rm sur}=0.2068$, as expected from Equation~(\ref{eq: surface_rad_flux_2}). The small reduction arises from the finite residual kinetic energy and gas enthalpy retained at the lower computational boundary. The numerical solution satisfies the vanishing net radiation-flux condition, $F_{\rm rad}(R_*)=0$, while the residual inflow velocity remains sufficiently small to represent approximate stagnation.

Figure~\ref{fig: NGC300} also shows the parallel scattering and free--free absorption optical depths, both integrated downward from the column top. The scattering optical depth increases monotonically through the sinking region and reaches $\tau_{\parallel}^{\rm scat}\simeq21.90$ at the stellar surface. The free--free absorption optical depth remains below unity through the upper part of the column and crosses unity at $r_{\rm th}=11.534\,{\rm km}$, thereby defining the thermal-mound surface. It subsequently rises to $\tau_{\parallel}^{\rm ff}\simeq4.94$ at the stellar surface. Thus, in the present model, the lower $\simeq1.5\,{\rm km}$ of the accretion column is optically thick to free--free absorption, with free--free processes becoming increasingly important.

For NGC 300 ULX-1, we adopt a surface magnetic-field strength of $B_*=6\times10^{12}\,{\rm G}$. This value is motivated by the rapid spin-period evolution of the source modeled by \citet{2018A&A...620L..12V} using standard disk--magnetosphere accretion-torque theory. \citet{2019MNRAS.488.5225V} showed that the inferred field can be lower when super-Eddington outflows are included. Nevertheless, $B_*=6\times10^{12}\,{\rm G}$ provides a representative torque-based estimate for the no-outflow interpretation and is therefore adopted as our fiducial value.

\subsection{Swift J0243.6$+$6124}
\label{sec: case_2}

\begin{figure*}[ht]
    \centering
    \includegraphics[width=15cm]{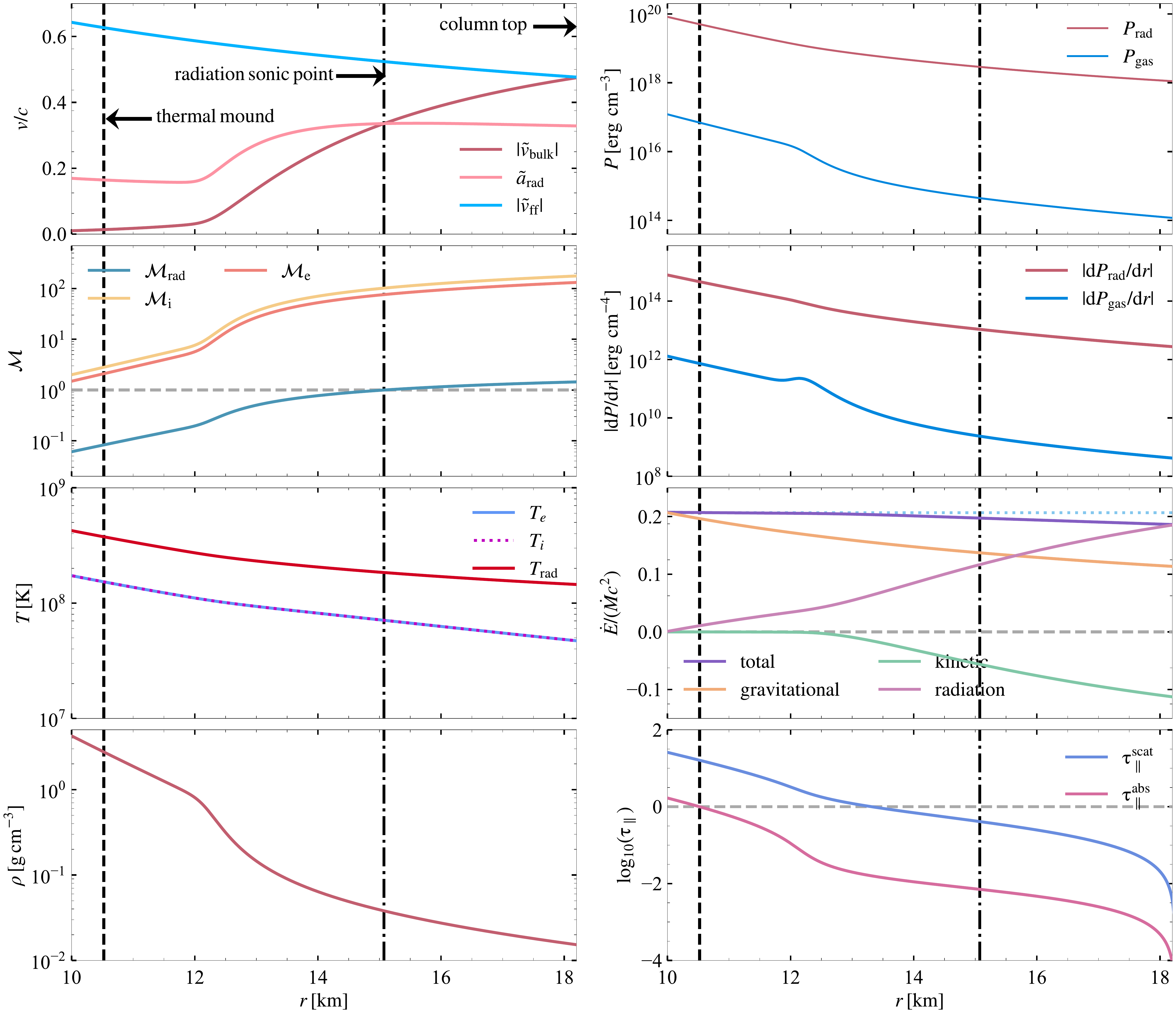}
    \caption{ Same as Figure~\ref{fig: NGC300}, but for the accretion column of Swift J0243.6$+$6124. The model is computed using the adopted model parameters and constrained source parameters listed in Tables~\ref{tab: model_parameters} and \ref{tab: constrained_parameters}, respectively.}
    \label{fig: 0243}
\end{figure*}

Figure~\ref{fig: 0243} shows the calculated accretion-column structure of Swift J0243.6$+$6124. We adopt an X-ray luminosity of $L_{\rm X}=2.0\times10^{39}\,{\rm erg\,s^{-1}}$, based on the earlier distance estimate of $D\simeq7\,{\rm kpc}$ \citep{2018ApJ...863....9W}, corresponding to a mass accretion rate of $\dot M=1.07\times10^{19}\,{\rm g\,s^{-1}}$ for the adopted stellar mass and radius. The updated Gaia EDR3 distance of $D=5.2\pm0.3\,{\rm kpc}$ implies a lower peak luminosity of approximately $1.1\times10^{39}\,{\rm erg\,s^{-1}}$ \citep{2021AJ....161..147B, 2026ApJ...998..184C}. However, since our primary aim is to investigate the radiation-hydrodynamical structure of accretion columns under representative super-Eddington conditions rather than to perform source-specific spectral fitting, we retain the earlier luminosity estimate as the fiducial input for the present calculation. The adopted model parameters are $\bar{\sigma}/\sigma_{\rm T}=5.50\times10^{-4}$, $\sigma_{\parallel}/\sigma_{\rm T}=1.37\times10^{-4}$, $\ell_1=550\,{\rm m}$, $\ell_2=750\,{\rm m}$, $M_{\rm rad,top}=1.45$, and $B_*=1.6\times10^{13}\,{\rm G}$. The resulting column top is located at $r_{\rm top}=18.251\,{\rm km}$, corresponding to a column height of $8.251\,{\rm km}$ above the stellar surface.

The radiation sonic surface, defined by $a_{\rm rad}=|v|$, is located at $r_{\rm sonic}=15.079\,{\rm km}$. Below this radius, the flow becomes subsonic with respect to radiation-pressure perturbations and enters the extended sinking regime. Radiation pressure progressively decelerates the inflow toward the stellar surface, while mass continuity causes the density to increase as the bulk velocity decreases. The radiation and gas pressures also rise toward the base of the column.

The thermal-mound surface is determined from the free--free thermalization condition $\tau_{\parallel}^{\rm ff}=1$. For the adopted Swift J0243.6$+$6124 parameters, we obtain $r_{\rm th}=10.532\,{\rm km}$, corresponding to a vertical extent $h_{\rm th} \equiv r_{\rm th}-R_* =0.532\,{\rm km}$. The electron temperature at the thermal-mound surface is $T_e(r_{\rm th})\simeq1.52\times10^{8}\,{\rm K}$. The thermal mound is therefore substantially more compact than that obtained for NGC 300 ULX-1, for which $h_{\rm th}=1.534\,{\rm km}$. This difference results from the combined effects of the lower mass accretion rate, the narrower accretion-column cross section, and the different magnetic and scattering parameters. Since the Rosseland mean free--free absorption coefficient scales approximately as $\alpha_R\propto n_e^2T_e^{-7/2}$, relatively modest variations in the near-surface density and temperature can produce appreciable shifts in the radius at which $\tau_{\parallel}^{\rm ff}=1$ is reached. The parallel optical-depth profiles provide an additional measure of the vertical structure of the column. The scattering optical depth increases steadily from the column top and reaches $\tau_{\parallel}^{\rm scat}\simeq26.14$ at the stellar surface. The free--free absorption optical depth remains below unity throughout most of the column and crosses unity only at $r_{\rm th}=10.532\,{\rm km}$. It subsequently increases to $\tau_{\parallel}^{\rm ff}\simeq1.70$ at the stellar surface. Hence, the region that is optically thick to free--free absorption is confined to the lowermost $\simeq0.53\,{\rm km}$ of the column, considerably narrower than in the NGC 300 ULX-1 model.

In contrast to NGC 300 ULX-1, for which the adopted magnetic-field
strength is motivated primarily by accretion-torque modeling,
Swift J0243.6$+$6124 has a more direct magnetic-field constraint from
cyclotron resonant scattering. During the 2017 giant outburst,
Insight-HXMT detected a high-energy CRSF at $E_{\rm cyc}\simeq120$--$146\,{\rm keV}$, implying a near-surface magnetic field of order $B_*\sim1.6\times10^{13}\,{\rm G}$ \citep{2022ApJ...933L...3K}. We therefore adopt $B_*=1.6\times10^{13}\,{\rm G}$ as the fiducial surface magnetic field in the present dynamical calculation.

\subsection{SMC X-3}
\label{sec: case_3}

\begin{figure*}[ht]
    \centering
    \includegraphics[width=15cm]{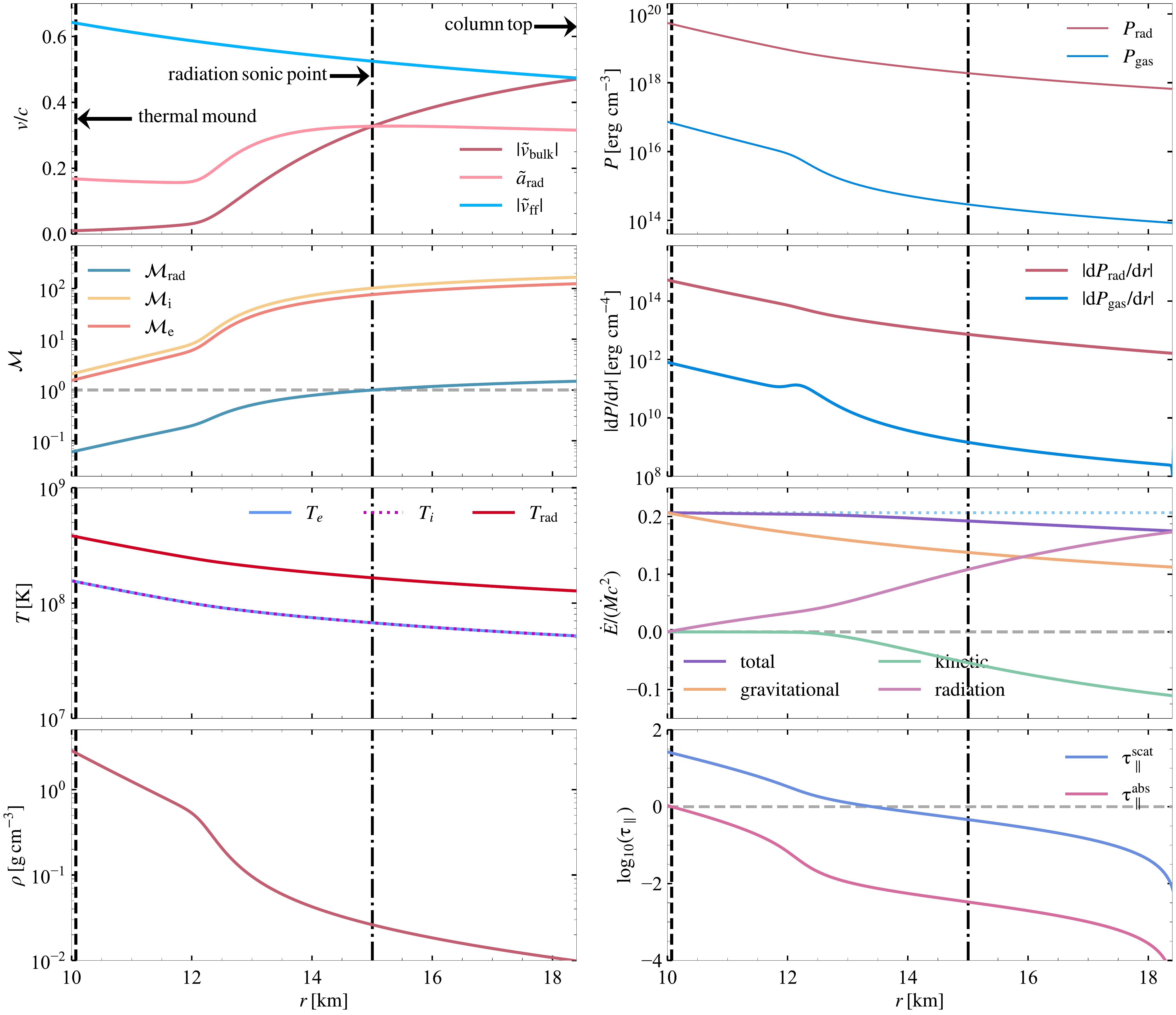}
    \caption{ Same as Figure~\ref{fig: NGC300}, but for the accretion
    column of SMC X-3. The model is computed using the adopted model parameters and constrained source parameters listed in Tables~\ref{tab: model_parameters} and \ref{tab: constrained_parameters}, respectively.}
    \label{fig: SMC}
\end{figure*}

Figure~\ref{fig: SMC} shows the calculated accretion-column structure
of SMC X-3. We adopt an X-ray luminosity of $L_{\rm X}=1.2\times10^{39}\,{\rm erg\,s^{-1}}$, corresponding to a mass
accretion rate of $\dot M=6.54\times10^{18}\,{\rm g\,s^{-1}}$ for the adopted stellar mass and radius. The adopted model parameters are
$\bar{\sigma}/\sigma_{\rm T}=4.00\times10^{-4}$,
$\sigma_{\parallel}/\sigma_{\rm T}=2.11\times10^{-4}$,
$\ell_1=500\,{\rm m}$, $\ell_2=700\,{\rm m}$, $\mathcal{M}_{\rm rad,top}=1.50$, and $B_*=4\times10^{13}\,{\rm G}$.  The resulting column top is located at $r_{\rm top}=18.494\,{\rm km}$, corresponding to a column height of $8.494\,{\rm km}$ above the stellar surface.

The radiation sonic surface, defined by $a_{\rm rad}=|v|$, is located at $r_{\rm sonic}=15.004\,{\rm km}$, comparable to the values obtained for NGC 300 ULX-1 and Swift J0243.6$+$6124. Below this radius, the inflow becomes subsonic with respect to radiation-pressure perturbations and enters an extended sinking regime. The radiation pressure gradient progressively decelerates the accreting material, while the density and pressure rise toward the stellar surface as required by mass continuity.

The most pronounced difference among the three sources is the structure of the thermal mound. For SMC X-3, the free--free thermalization condition,
$\tau_{\parallel}^{\rm ff}=1$, is reached at $r_{\rm th}=10.070\,{\rm km}$, corresponding to a vertical extent $h_{\rm th} \equiv r_{\rm th}-R_* =0.070\,{\rm km}$. The electron temperature at the thermal-mound surface is $T_e(r_{\rm th})\simeq1.53\times10^{8}\,{\rm K}$. The mound is therefore much more compact than those obtained for NGC 300 ULX-1 and Swift J0243.6$+$6124, for which $h_{\rm th}=1.534\,{\rm km}$ and $0.532\,{\rm km}$, respectively. This difference reflects the strong sensitivity of the free--free absorption coefficient to the local density and electron temperature,
\begin{equation}
\alpha_R\propto n_e^2T_e^{-7/2} \,.
\end{equation}
Hence, relatively modest changes in the near-surface thermodynamic profiles can produce substantial shifts in the radius at which $\tau_{\parallel}^{\rm ff}=1$ is reached.

The compact thermal mound in SMC X-3 is also evident from the optical depths. The parallel scattering optical depth increases monotonically from the column top and reaches $\tau_{\parallel}^{\rm scat}\simeq26.69$ at the stellar surface. By contrast, the free--free absorption optical depth remains below unity through almost the entire accretion column. It crosses unity only at $r_{\rm th}=10.070\,{\rm km}$ and rises to merely $\tau_{\parallel}^{\rm ff}\simeq1.071$ at the stellar surface. Thus, the region that is optically thick to free--free absorption is confined to the lowermost $\simeq70\,{\rm m}$ of the column. This is considerably thinner than the corresponding regions in the NGC 300 ULX-1 and Swift J0243.6$+$6124 models.

The adopted magnetic-field strength is motivated by the pulse-profile transition observed during the 2016 giant outburst of SMC X-3. \citet{2022MNRAS.517.3354L} reported a phase shift of approximately $0.25$ between the high- and low-luminosity double-peaked pulse profiles, which they interpreted as evidence for a transition between fan-beam and pencil-beam emission. Using the corresponding critical luminosity, they inferred a surface magnetic-field strength of order $B_*\sim4\times10^{13}\,{\rm G}$. We therefore adopt $B_*=4\times10^{13}\,{\rm G}$ as the fiducial magnetic field in the present calculation. Together with the adopted parallel scattering cross section and polar-cap geometry, this field strength yields a radiation-dominated sinking flow and a very compact near-surface thermalization layer.

\section{Discussion}
\label{sec: discussion}

In this work, we have applied the two-fluid radiation-hydrodynamical framework of \citet{2017ApJ...835..129W} to investigate the accretion-column structures of NGC 300 ULX-1, Swift J0243.6$+$6124, and SMC X-3 under representative super-Eddington accretion conditions. Following the original formulation, we solve the coupled conservation equations in a dipole-funnel geometry, retaining separate electron and ion energy equations while adopting a simplified Compton energy-exchange prescription without explicitly solving the photon-transport equation. Using observationally motivated accretion rates, magnetic-field strengths, and adopted column geometries, we calculate and compare the radial profiles of velocity, density, temperature, pressure, and energy transport, with particular emphasis on the locations of the radiation sonic surfaces and thermal mounds.

A key difference between the ULXPs studied here and the luminous X-ray pulsars Her X-1, Cen X-3, and LMC X-4 is the much larger mass accretion rate. In the ULXP regime, the enhanced radiation pressure strongly modifies the structure of the column. In particular, the energy transport near the column top can already be dominated by radiation rather than by kinetic energy, unlike the lower-luminosity pulsars modeled by \citet{2017ApJ...835..129W}. This indicates that ULXP columns are not simply scaled-up versions of ordinary bright X-ray pulsar columns; instead, their larger accretion rates lead to a more extended radiation-dominated deceleration region.

All three sources studied here develop an extended sinking region above the stellar surface. In this region, the inflowing gas is gradually decelerated by the radiation pressure gradient, while the density, temperature, gas pressure, and radiation pressure increase toward the base of the column. The radiation sonic surfaces are located at $r_{\rm sonic}=15.032$, $15.079$, and $15.004\,{\rm km}$ for NGC 300 ULX-1, Swift J0243.6$+$6124, and SMC X-3, respectively. Below this surface, the flow becomes subsonic with respect to radiation-pressure perturbations, and the energy transport is dominated by radiation rather than by kinetic-energy transport.

The thermal mound forms close to the stellar surface in all three systems. Its location is determined by the parallel Rosseland mean free--free optical depth, rather than imposed as an arbitrary inner boundary. The resulting mound heights are $1.534\,{\rm km}$ for NGC 300 ULX-1, $0.532\,{\rm km}$ for Swift J0243.6$+$6124, and $0.070\,{\rm km}$ for SMC X-3. The much smaller mound height in SMC X-3 shows that the thermalization layer in this source is more tightly confined to the stellar surface. This difference reflects the sensitivity of the free--free absorption depth to the near-surface density, temperature, accretion rate, column geometry, and effective scattering opacity.

A similar location for the radiation sonic surface was found by \citet{2024ApJ...966L...5W} in their modeling of RX J0209.6$-$7427. They obtained $r_{\rm sonic}=14.91$ and $15.34\,{\rm km}$ in the low- and high-luminosity states, respectively. These values are comparable to those obtained for the three ULXPs considered here. This similarity suggests that an extended radiation-dominated deceleration region may be a common feature of super-Eddington accretion columns. However, a broader exploration of the model parameters is required to establish the generality of this result.

The mass accretion rate is a key input parameter in radiation-hydrodynamical calculations of accretion columns. An important distinction between ULXPs and lower-luminosity X-ray pulsars, such as Her X-1, LMC X-4, and Cen X-3 \citep{2007ApJ...654..435B}, is that the high mass-transfer rates associated with ULXPs may drive the inner accretion disk into the supercritical regime. When the spherization radius exceeds the magnetospheric radius, $R_{\rm sp}>R_m$, radiation-driven winds can remove a substantial fraction of the inflowing material, while advective energy transport modifies the energy available to power these outflows \citep{1973A&A....24..337S, 1999AstL...25..508L, 2007MNRAS.377.1187P, 2019MNRAS.484..687M}. Consequently, the mass accretion rate reaching the magnetosphere, $\dot M(R_m)$, may differ significantly from the mass-transfer rate supplied by the companion, $\dot M_0$.

In the present work, we estimate the effective accretion rate from the observed X-ray luminosity using Equation~(\ref{eq: mdot_lx}), assuming that the emission is powered predominantly by accretion onto the neutron-star surface. This prescription can be applied to both lower-luminosity X-ray pulsars and ULXPs under the same assumptions concerning radiative efficiency and emission anisotropy. However, when estimating the accretion rate from the donor mass-transfer rate, the effects of supercritical disk winds and advection must be considered whenever $R_{\rm sp}>R_m$. To account for this distinction, we summarize a supercritical disk model in Appendix~A, which provides a framework for relating $\dot M_0$ to the accretion rate supplied to the magnetosphere, $\dot M(R_m)$.

Several limitations of the present calculation should be kept in mind. First, the model is cross-section averaged and does not resolve multidimensional structures or instabilities. Second, the magnetic scattering cross sections are treated as effective energy-averaged parameters. Third, the inverse-Compton temperature is not computed from the photon distribution function. The net Compton heating and cooling term is therefore neglected in the present hydrodynamical solutions. These approximations allow us to isolate the vertical dynamical structure of the accretion column, but future work should couple the hydrodynamical profiles to a full photon-transport calculation.

\section{Conclusion}
\label{sec: conclusion}

We have investigated the vertical dynamical structures of the accretion columns in NGC 300 ULX-1, Swift J0243.6$+$6124, and SMC X-3 using the two-fluid radiation-hydrodynamical framework of \citet{2017ApJ...835..129W}. By applying the coupled conservation equations to three ULXPs with different accretion rates, magnetic fields, and column geometries, we obtain and compare their velocity, density, temperature, pressure, and energy-transport profiles. These calculations provide a comparative description of radiation-dominated accretion columns under representative super-Eddington conditions.

Our main result is that all three ULXPs develop extended radiation-dominated sinking regions above the stellar surface. The radiation sonic surfaces are located at 15.032, 15.079, and 15.004 km for NGC 300 ULX-1, Swift J0243.6$+$6124, and SMC X-3, respectively, indicating that radiation pressure controls the flow over a large fraction of the column. Below the sonic surface, the inflowing matter is progressively decelerated, while the density and pressure increase toward the column base.

The thermal-mound location is determined from the condition that the parallel free--free absorption optical depth reaches unity. The resulting mound heights are 1.534, 0.532, and 0.070~km for NGC 300 ULX-1, Swift J0243.6$+$6124, and SMC X-3, respectively. These results show that the thermalization layer is confined to the lower part of the column and is particularly compact in SMC X-3. In all three sources, the residual velocity at the stellar surface is only about $0.01c$, indicating that most of the inflow has been decelerated before reaching the neutron-star surface.

The dynamical profiles obtained here provide a physical basis for interpreting radiation-dominated accretion columns in ULXPs. A natural next step is to couple the hydrodynamical solutions to a full photon-transport calculation in order to determine the emergent spectrum, inverse-Compton temperature, and energy-dependent radiation pattern self-consistently. Future high-sensitivity X-ray timing, spectroscopic, and polarimetric observations, including those expected from eXTP, will provide new opportunities to test this picture and constrain the geometry and magnetic-field properties of super-Eddington pulsars \citep{Zhang:2025iae, Ge:2025upe, Li:2025uaw, Bu:2025jxy, Zhou:2025pth}.

\begin{acknowledgments}

This work was supported by the National Natural Science Foundation of China (Grant No. 8200910057, 12447148, 12247101), the Fundamental Research Funds for the Central Universities (Grant No. lzujbky-2026-jdzx02), the Natural Science Foundation of Gansu Province (No. 25JRRA799), the ‘111 Center’ under Grant No. B20063, the National SKA Program of China (2020SKA0120100), and the High-Performance Computing Platform of Peking University.

\end{acknowledgments}

\appendix

\section{Mass Accretion Rates in Standard NSULXs}
\label{sec: accretion_rate}

The high mass accretion rates associated with ULXPs may drive their inner accretion disks into the supercritical regime, distinguishing their accretion environments from those of lower-luminosity X-ray pulsars, such as Her X-1 and Cen X-3. When the spherization radius exceeds the magnetospheric radius, radiation-driven winds can remove part of the transferred material, while advective energy transport modifies the energy available to drive the outflows \citep{1973A&A....24..337S, 1999AstL...25..508L,
2007MNRAS.377.1187P, 2019MNRAS.484..687M}.

The radiation-hydrodynamical equations solved in this work require the mass accretion rate entering the magnetic accretion column as an input parameter. In the main text, we estimate this quantity from the observed X-ray luminosity, assuming that the radiation is predominantly powered by accretion onto the neutron-star surface. An alternative approach is to determine the mass accretion rate reaching the magnetosphere from the mass-transfer rate supplied by the companion, explicitly accounting for mass loss through supercritical disk winds.

In this Appendix, we summarize a supercritical disk prescription that relates the donor mass-transfer rate, $\dot M_0$, to the accretion rate reaching the magnetosphere, $\dot M(R_m)$. This framework allows us to assess how disk outflows and advective transport modify the mass supply to the accretion column under supercritical accretion conditions.

For a magnetized neutron star, we identify the inner disc radius with
the magnetospheric truncation radius,
$R_{\rm in}\equiv R_{\rm m}$. Following
\citet{2019MNRAS.484..687M}, the magnetospheric radius is approximated
as
\begin{equation}
R_{\rm m}
\simeq
5.6\times10^7
\Lambda
B_{{\rm d},12}^{4/7}
\dot M_{19}^{-2/7}
M_{1.4}^{-1/7}
R_6^{12/7}
\ {\rm cm},
\label{eq:magnetospheric_radius_appendix}
\end{equation}
where
$B_{{\rm d},12}=B_{\rm d,*}/10^{12}\,{\rm G}$ is the surface strength
of the large-scale dipole magnetic field, $\dot M_{19}=\dot M(R_{\rm m})/ (10^{19}\,{\rm g\,s^{-1}})$, $M_{1.4}=M_*/(1.4M_\odot)$, and $R_6=R_*/10^6\,{\rm cm}$.  The dimensionless factor $\Lambda$ accounts for uncertainties in the accretion geometry and disc--magnetosphere coupling. Following the commonly adopted prescription for disc accretion onto magnetized neutron stars, we set $\Lambda=0.5$ in the present calculations.

At super-Eddington mass-transfer rates, the disc becomes geometrically
thick inside the spherization radius. We assume that a fraction
$\epsilon_{\rm w}$ of the locally dissipated energy is used to power
radiation-driven winds, with $0\leq\epsilon_{\rm w}\leq1$. Following
\citet{1973A&A....24..337S,1999AstL...25..508L,
2007MNRAS.377.1187P}, the spherization radius is approximated as
\begin{equation}\label{eq: R_sp}
\begin{split}
R_{\rm sp} \approx\
9\times10^5\,\dot m
\bigg[
&1.34 - 0.4\epsilon_{\rm w}
+0.1\epsilon_{\rm w}^2
\\
&-\left(1.1-0.7\epsilon_{\rm w}\right)
\dot m^{-2/3}
\bigg]
\,{\rm cm},
\end{split}
\end{equation}
where
\begin{equation}
\dot m
\equiv
\frac{\dot M_0}{\dot M_{\rm Edd}}.
\end{equation}
For consistency with the supercritical-disc prescription adopted here,
we use the Eddington mass-accretion-rate convention
\begin{equation}
\dot M_{\rm Edd}
\simeq
1.9\times10^{18}
R_6
\ {\rm g\,s^{-1}}.
\label{eq:mdot_edd_appendix}
\end{equation}
We denote the auxiliary mass accretion rate at the innermost stable circular orbit (ISCO) by $\dot M_{\rm ISCO}$.

Advective transport carries part of the viscously dissipated energy
inward, thereby reducing the fraction of the released energy available
to drive the wind. Following \citet{2007MNRAS.377.1187P} and
\citet{2019MNRAS.484..687M}, the mass accretion rate at a radial
coordinate $r<R_{\rm sp}$ is approximated by
\begin{equation}\label{eq: Mdot_r}
\dot M(r)
=
\dot M_{\rm ISCO}
+
\left(
\dot M_0-\dot M_{\rm ISCO}
\right)
\frac{r}{R_{\rm sp}}\,.
\end{equation}
The auxiliary quantity $\dot M_{\rm ISCO}$ is given by
\begin{equation}\label{eq: Mdot_ISCO}
\dot M_{\rm ISCO}
=
\dot M_0
\frac{1-\mathcal A_{\rm w}}
{1-\mathcal A_{\rm w}
(0.4\,\dot m)^{-1/2}}\,,
\end{equation}
where
\begin{equation}\label{eq:Aw}
\mathcal A_{\rm w}
\equiv
\epsilon_{\rm w}
\left(
0.83-0.25\epsilon_{\rm w}
\right) \,.
\end{equation}
The notation $\dot M_{\rm ISCO}$ follows \citet{2019MNRAS.484..687M}, based on the supercritical-disc prescription of \citet{2007MNRAS.377.1187P}. For a magnetized neutron star, however, the disc is generally truncated at the magnetospheric radius before reaching the ISCO, and therefore $\dot M_{\rm ISCO}$ should be regarded as an auxiliary quantity rather than the physical mass accretion rate entering the magnetosphere.

The physically relevant accretion rate feeding the magnetic
accretion column is $\dot M(R_{\rm m})$. When
$R_{\rm m}<R_{\rm sp}$, the magnetosphere lies inside the
wind-producing region, and Equations~
(\ref{eq:magnetospheric_radius_appendix}) and
(\ref{eq: Mdot_r}) must be solved simultaneously because
$R_{\rm m}$ depends on $\dot M(R_{\rm m})$, while the latter depends
on the position of $R_{\rm m}$. In this case,
\begin{equation}\label{eq:Mdot_Rm}
\dot M(R_{\rm m})
=
\dot M_{\rm ISCO}
+
\left(
\dot M_0-\dot M_{\rm ISCO}
\right)
\frac{R_{\rm m}}{R_{\rm sp}} \,.
\end{equation}
If instead $R_{\rm m}\geq R_{\rm sp}$, the disc is truncated before
entering the wind-producing region, and we approximate
\begin{equation}
\dot M(R_{\rm m})
=
\dot M_0 \,.
\end{equation}

\begin{figure}
    \centering
    \includegraphics[width=8cm]{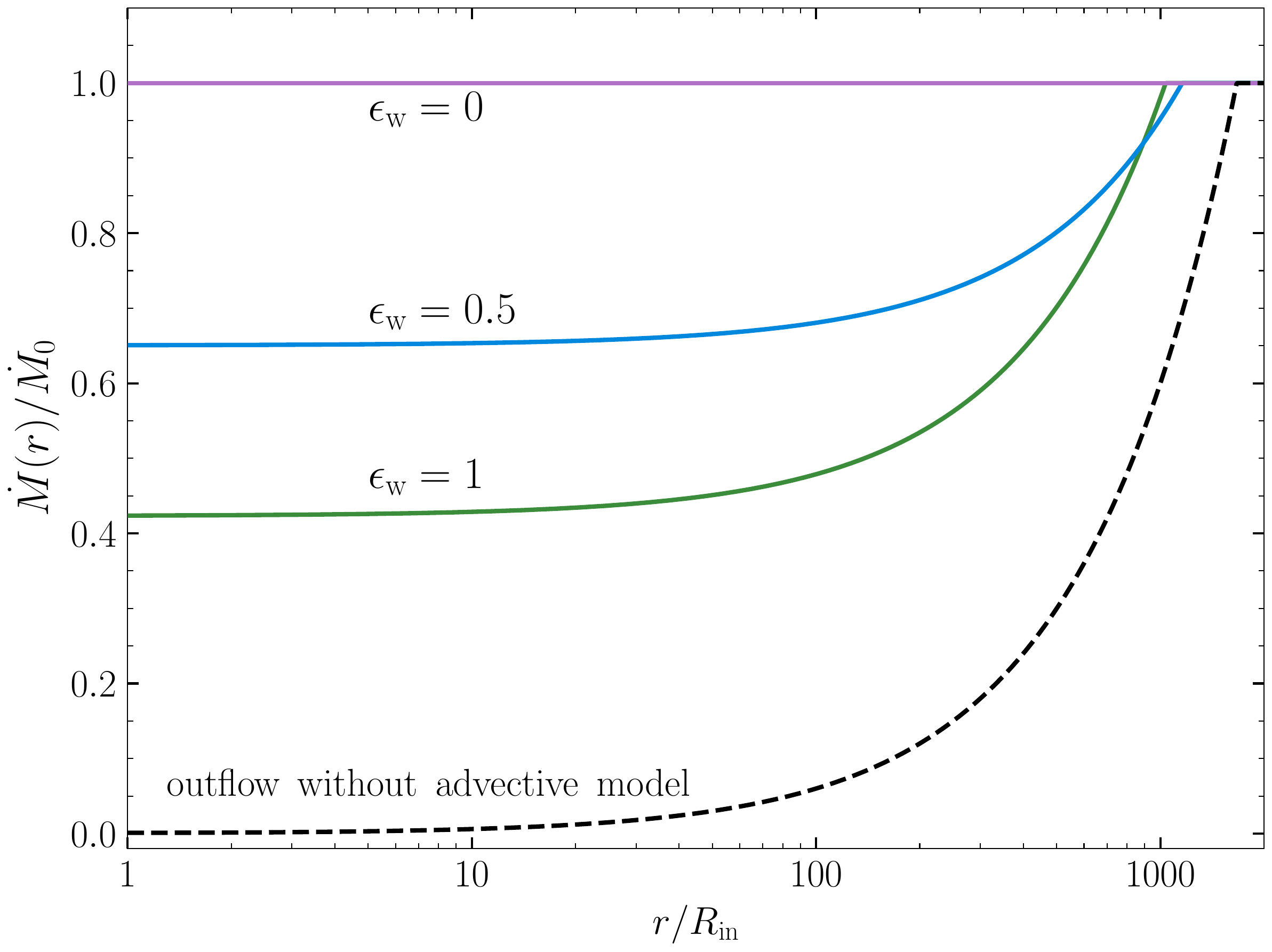}
    \caption{
    Radial profile of the normalized mass accretion rate for an
    initial dimensionless mass-transfer rate $\dot m=10^3$.
    The solid curves include wind-driven mass loss and advective
    transport of viscously generated heat for different values of the
    wind efficiency $\epsilon_{\rm w}$. The dashed line shows the
    corresponding non-advective solution with outflows
    \citep{1999AstL...25..508L,2007MNRAS.377.1187P}.
    }
    \label{fig: accretion_rate}
\end{figure}

Figure~\ref{fig: accretion_rate} shows an illustrative radial profile of the normalized mass accretion rate for an initial dimensionless mass-transfer rate $\dot m=10^3$. The curves include wind-driven mass loss together with advective transport of viscously generated heat. Because advection carries part of the dissipated energy inward, only a fraction of the donor-supplied material is removed by the wind, while the remainder can propagate inward and potentially reach the magnetosphere. The resulting $\dot M(R_{\rm m})$ depends primarily on the wind efficiency $\epsilon_{\rm w}$, the donor mass-transfer rate $\dot M_0$, and the surface dipole magnetic field $B_{\rm d,*}$, with $\Lambda$ fixed at $0.5$.

Once these quantities are specified, Equations~
(\ref{eq:magnetospheric_radius_appendix})--(\ref{eq:Mdot_Rm}) provide
an estimate of both the magnetospheric radius and the mass accretion
rate ultimately supplied to the magnetosphere. For a main-sequence
donor transferring mass on its thermal timescale, the characteristic
mass-transfer rate can be approximated as
\citep{1999ApJ...519L.169K,1999MNRAS.309..253K}
\begin{equation}\label{eq:thermal_mass_transfer}
\dot M_0
\simeq
3\times10^{-8}
\left(
\frac{M_2}{M_\odot}
\right)^{2.8}
\,M_\odot\,{\rm yr^{-1}} \,,
\end{equation}
where $M_2$ is the donor mass. Together with the supercritical-disc
prescription described above, this relation provides a connection
between binary mass transfer and the mass accretion rate entering the
magnetosphere, which is the appropriate input quantity for the
radiation-hydrodynamical accretion-column model.

The treatment presented in this Appendix is intended to provide a
framework for connecting the mass-transfer rate in a binary system to
the mass accretion rate supplied to the magnetic accretion column.
A detailed source-by-source application to standard NSULXs, including
constraints on $\epsilon_{\rm w}$, $\Lambda$, and the dipole magnetic
field, will be presented elsewhere.

\bibliography{Model_ULXPs}{}
\bibliographystyle{aasjournalv7}

\end{document}